\documentclass[10pt]{iopart} % For two-column output uncomment the next line and choose [10pt] rather than [12pt] in the \documentclass declaration
\ioptwocol

\expandafter\let\csname equation*\endcsname\relax
\expandafter\let\csname endequation*\endcsname\relax
\usepackage{amsmath}

\usepackage{iopams}
\usepackage{natbib}
\usepackage{hyperref}
\usepackage{aas_macros}
\usepackage{orcidlink}
\usepackage{enumitem}
\usepackage{pdflscape}
\usepackage{booktabs}
\usepackage{xspace}

\usepackage{subcaption} % for subfigures

\newcommand{\dd}{\ensuremath{\mathrm{d}}}

\newcommand{\Mchirp}{\ensuremath{\,{\rm M}_{\rm c}}\xspace}
\newcommand{\chieff}{\ensuremath{\,\chi_{\rm eff}}\xspace}

\newcommand{\Msun}{\ensuremath{\,\mathrm{M}_{\odot}}}

\newcommand{\Mtr}{\ensuremath{\,\rm{M}_{\rm tr}}}
\newcommand{\MBHmax}{\ensuremath{\,\rm{M}_{\rm BH,max}}}

\newcommand{\edit}[1]{{\textbf{\color{red}#1}}}
\renewcommand{\edit}[1]{#1}

\usepackage[deletedmarkup=xout,commentmarkup=uwave,commandnameprefix=ifneeded]{changes}
\definechangesauthor[name=Soumendra, color=teal]{SR}
\definechangesauthor[name=Lieke, color=orange]{LvS}
\definechangesauthor[name=Will, color=magenta]{WF}
\usepackage{xcolor}

\graphicspath{ {./figures/} }

\begin{document}

\title[Mid-Thirties Crisis]{A Mid-Thirties Crisis: Dissecting the Properties of Gravitational Wave Sources Near the 35 Solar Mass Peak}

\author{Soumendra Kishore Roy\orcidlink{0000-0001-9295-5119}$^1$,
        Lieke A. C. van Son\orcidlink{0000-0001-5484-4987}$^{2,3,4}$,
        Will M. Farr\orcidlink{0000-0003-1540-8562}$^{1,2}$}

\address{$^1$Department of Physics and Astronomy, Stony Brook University, Stony Brook, NY 11794, USA}
\address{$^2$Center for Computational Astrophysics, Flatiron Institute, 162 Fifth Avenue, New York, NY 10010, USA}
\address{$^3$Department of Astrophysical Sciences, Princeton University, 4 Ivy Lane, Princeton, NJ 08544, USA}
\address{$^4$Department of Astrophysics/IMAPP, Radboud University, P.O. Box 9010, NL-6500 GL Nijmegen, The Netherlands}

\ead{soumendrakisho.roy@stonybrook.edu}

\vspace{10pt}

\begin{indented}
\item[]Date: \today
\end{indented}

\begin{abstract}
% intro
One striking feature of binary black hole (BBH) mergers observed in the first decade of gravitational-wave astronomy is an excess of events with component masses around $35\Msun$. Multiple formation channels have been proposed to explain this excess. To distinguish among these channels, it is essential to examine their predicted population-level distributions across additional parameters.
% What we do
In this work, we focus on BBH mergers near the $35\Msun$ peak and infer the population distributions of primary mass ($m_1$), mass ratio ($q$), effective spin ($\chieff$), and redshift ($z$).
% What we find
We observe a gradual increase in the merger rate with $m_1$, rising by a factor of $3$ from $20\Msun$ to a peak around $34\Msun$, followed by a sharp, order-of-magnitude decline by $50\Msun$. This population also shows a weak preference for equal-mass mergers and has a $\chieff$ distribution skewed toward positive values, with a median of zero excluded at approximately $90\%$ confidence. We find no significant $q-\chieff$ correlation in the $35 \Msun$ peak population, suggesting that lower-mass systems ($m_1<20\Msun$) likely drive the $q-\chieff$ anti-correlation observed in the full BBH merger catalog.
The redshift evolution of the merger rate is consistent with the cosmic star formation rate.
% comparison to astro
We compare our findings with predictions from a wide range of formation channels. We find that common variants of the pair-instability supernova scenario, as well as hierarchical mergers \edit{in absence of sufficient gas-accretion}, are incompatible with the observed features of the $35\Msun$ population.
% Some outlooking thoughts
Ultimately, none of the formation channels we consider can explain all or even most of the features observed in this population. The ``mid-thirties'' of black hole mergers are in crisis.
\end{abstract}

%\maketitle

% \section{leaving comments}
% Here's how you leave comments:  \chcomment[id=LvS]{Lieke stole this from Max from another paper} 
% Suggest new text \added[id=LvS]{This is some added text.}, or replace text \replaced[id=LvS]{new text}{old text}
% You can also mark text for deletion: \deleted[id=LvS]{text I don't like}.
% The \texttt{changes} package has a \texttt{final} import option that will show you what the PDF looks like without all the comments (accepting all suggested text). 

%%%%%%%%%%%%%%%%%%%%%%%%%%%%%%%%%%%%%%%%%%%%%%%%%%%%%%%%%%%%%%%%%%%%%%%%%%%%%%%%%%%%%%%%%%%%%%%%%%%%%%
\section{Introduction}\label{sec:Intro}

% Birthday of 35Msun BBHs!
It has been a full decade since the first detection of a binary black hole (BBH) merger, GW150914 \citep{GW150914}. 
This groundbreaking detection revealed component masses of $m_1 = 36^{+5}_{-4}\Msun$, and $m_2 = 29^{+4}_{-4}\Msun$, where $m_1$ and $m_2$ represent the primary (heavier) and secondary (lighter) black-hole masses, respectively.
These masses were unexpectedly high, exceeding the most massive stellar-mass black holes known at the time \citep{Tetarenko2016}. % Cyg X-1 at the time was though to be 14.8±1.0M \cite{Orosz2011})
Yet, as the number of detections grew, it became clear that merging BBHs with primary masses around $35\Msun$ are not outliers.
In fact, most of the observed gravitational-wave (GW) systems detected since then have a primary mass between $30$ and $40\Msun$\citep{GWTC3Catalog, Nitz_2019, Nitz_2020, Nitz_2021, Nitz_2023, Zackay_2019, Venumadhav_2020, Olsen_2022, Mehta_2025}, creating a noticeable excess near $35\Msun$ \citep{gwtc-3pop}.

%%%%%%%%%%%%%%%%%%%%%%%%%%%%%%%%%%%%%%%%%%%%%%%%%%%
% History of the 35 Msun peak model
\paragraph{Modeling the $35\Msun$ Peak: A Brief History}
This excess has drawn much attention since the earliest detections. With just five confirmed events, \cite{FishbachHolz2017} noted a possible deficit of black holes (BHs) above $\sim40\Msun$, which was linked to the maximum BH mass as set by pair-instability in stellar cores \citep{FowlerHoyle1964,BarkatRakavy1967,RakavyShaviv1967,Fraley1968,Bond1984}. 
This led to new parameterized models for the BBH mass function, including a modelled excess of BHs before the maximum mass \citep{Kovetz2017,TalbotThrane2018, GWTC-2_BBH_prop2019}. It provides the basis for the `Power Law + Peak' model, which has since become the default model to describe the BBH mass function \citep{GWTC-2pop2021, gwtc-3pop}.
Since then, a growing body of evidence has supported the presence of a `bump' in the BBH mass function around $\sim35\Msun$.
This has been consistently confirmed through various models, including parametric \citep[typically mixtures of power laws and/or Gaussian peaks][]{gwtc-3pop,Toubiana2023,MaganaHernandez2024,Gennari2025}, 
more flexible `semi-parametric' models like cubic-spline perturbations \citep{Edelman2022}, and the `Power Law + Spline' model \citep{gwtc-3pop}, 
as well as many non-parametric approaches including piecewise-constant binned models \citep{gppop1, Fishbach_2020, Veske2021, gppop2, gwtc-3pop, Ray_2023, PhysRevD.111.L061305}, Gaussian mixture models \citep{Tiwari2021,TiwariFairhurst2021,gwtc-3pop,Rinaldi_2024}, methods based on adaptive kernel density estimates \citep{Sadiq2022,Sadiq_2023, gwtc-3pop}, basis splines \citep{Edelman2023,gwtc-3pop}, and autoregressive processes \citep{CallisterFarr2024}.

This extensive body of work has established this feature as the most statistically robust structure in the current GW catalog \citep[cf.][]{Farah2023}. 
However, while the statistical significance of this peak has grown, its astrophysical origin remains uncertain.

%%%%%%%%%%%%%%%%%%%%%%%%%%%%%%%%%%%%%%%%%%%%%%%%%%%
% Astro interpretation
\paragraph{Astrophysical origin of the $35 \Msun$ peak}

%%%%%%%%%%%%%%%%%%
% PPISN: why would it create a bump at 35Msun?
Initially, the prevailing explanation linked the $35\Msun$ feature to a pile-up in the remnant mass distribution caused by `pair-instability supernova' \citep[see e.g.,][]{TalbotThrane2018}.
% Stars with sufficiently massive cores reach temperatures and densities in their centers that allow for electron-pair production, triggering premature collapse of the carbon-oxygen core, resulting in an explosion that disintegrates the star without leaving a BH remnant \citep{BarkatRakavy1967,RakavyShaviv1967,Fraley1968,Bond1984,Woosley2017,RenzoSmith2024}.
This theory predicts a gap in the BH mass function between approximately $45-80\Msun$ and $135-160\Msun$, known as the pair-instability supernova (PISN) mass gap  \citep[see e.g.,][and references \edit{therein}]{Woosley_2021,RenzoSmith2024}.
Stars just below the PISN mass gap undergo pulsations (pulsational pair-instability supernova, P-PISN) that do not fully disrupt them but instead map a range of stellar masses to similar BH masses, predicting a pile-up just below the gap's lower bound \citep[e.g.,][]{Belczynski_PISNbbh2016,Woosley2017,SperaMapelli2017,Stevenson2019,Marchant2019,Karathanasis2023}.
%%%%%%%%%%%%%%%%%%%%
% Why PPISN does not work
While the occurrence of pair-instability is remarkably robust against most uncertainties in stellar evolution \citep{Takahashi2018,Renzo2020_conv_edge_ppisn,MarchantMoriya2020}, its \textit{location} is highly sensitive to the $\rm ^{12}C(\alpha,\gamma)^{16}O$ reaction rate \citep{Farmer_2019,Farmer2020}.

The location of the observed over-density is in the vicinity of the predicted lower bound of the PISN mass gap, but recent studies consistently place this boundary at \edit{a} notably higher mass of $\sim60^{+32}_{-14}\Msun$ \citep{Mehta2022,Farag2022,Shen2023}.
This places the P-PISN gap at odds with the observed $35\Msun$ feature \citep[][]{Farag2022,Hendriks2023,Golomb2024}.
Moreover, shifting the PISN gap to align with the GW-detected peak creates a tension with the observed rate of hydrogen-less super-luminous supernovae \citep{Hendriks2023}.
The true location of the lower edge of the PISN mass gap, and whether it has been observed, remains an open question \citep[though see][]{Li_2024,Ulrich2024,Antonini2025}.

%%%%%%%%%%%%%%%%%%%%
% Other proposed formation channels
Several alternative explanations for the $35\Msun$ peak have been proposed, including a signature of dynamical formation in globular clusters \citep[e.g.,][]{Antonini_2023,Ray2024}, population III \edit{(Pop III)} stars \citep{Kinugawa_2014, Kinugawa_2020, Kinugawa_2021}, a lower-mass shoulder of PISN \citep{CroonSakstein2023}, chemically homogeneous evolution (CHE, e.g. de Sa et al., in preparation), or the result of stable mass transfer and quasi-homogeneous evolution \citep{Briel2023}.  
Each scenario makes distinct predictions about the BBH merger population beyond `just' the masses, which can be tested against observations to identify the most likely formation pathway (see Section \ref{sec:astrointerp}).

%%%%%%%%%%%%%%%%%%%%%%%%%%%%%%%%%%%%%%%%%%%%%%%%%%%
\paragraph{Motivation for this work}
In this paper, we isolate BBH mergers near the $35 \Msun$ peak and analyze them separately from the rest of the population (see \edit{Section~\ref{sec:Method}} for details). 
This allows us to examine the multidimensional properties of BBH mergers that contribute to the $35\Msun$ feature. 
Specifically we aim to address the following questions for systems in this peak population:
\begin{enumerate}[label=\arabic*.]
    \item What is the \textit{shape} of their mass function?
 
    \item What is their mass ratio distribution?
    %How many equal- and un-equal-mass meregrs occur?

    \item What is their effective spin distribution?

    \item Does their effective spin distribution vary with mass ratio?

    \item How does their merger rate evolve with redshift?
\end{enumerate}

% To investigate these questions, 
For this purpose, we use the third Gravitational-Wave Transient
Catalog \citep[GWTC-3;][]{Abbott_2023}, published by the LIGO-Virgo-KAGRA
Collaboration \citep{aligo, avirgo, 10.1093/ptep/ptaa125}.
The rest of this paper is structured as follows. Section \ref{sec:Method} describes our population model and hierarchical inference framework, including selecting BBH mergers within a specific parameter range. Section \ref{sec:Result} presents our results. 
We compare the observed population with predictions from various astrophysical formation
channels in Section \ref{sec:astrointerp}.
% addresses their astrophysical interpretation. 
This reveals a ``mid-thirties crisis'' that we summarize in Section \ref{sec:conclusion} \edit{of our conclusion}.
% We conclude with a summary in 

%%%%%%%%%%%%%%%%%%%%%%%%%%%%%%%%%%%%%%%%%%%%%%%%%%%%%%%%%%%%%%%%%%%%%%%%%%%%%%%%%%%%%%%%%%%%%%%%%%%%%%
\section{Method}\label{sec:Method}

This section describes how we isolate BBH mergers near the $35 \, \Msun$ peak,
and study the multidimensional distribution of primary mass, mass ratio, spins,
and redshift. An alternative approach would be to apply a specialized model for
events near the peak while simultaneously using a separate, more flexible model
to describe the rest of the population. The latter would need the entire dataset
to be analyzed without explicitly isolating a subset of events \edit{\citep[for example, see][]{godfrey2023}}. Intrinsically, such an analysis would depend on how well
the more flexible model performs, and how tightly coupled it is to the dedicated
population part \footnote{Another alternative can be to compute the overlap between the parameter estimation
posterior and the expected \emph{detected} population from an astrophysical formation 
channel \citep[see][]{afroz2024phasespacebinaryblack}.}.

% \subsection{How do we select BBH mergers within a certain mass range?}
\subsection{Selecting BBH mergers with certain masses}\label{subsec:howcut}

The
standard methodology for dealing with event selection \citep{popgw1} requires
that the selection is based on data, not latent variables; thus it is not straightforward to select a subset of events corresponding to a region of validity in the latent parameter space of a particular formation channel or
population feature.

\begin{figure*}%[h!]
    \centering

    \includegraphics[width=0.65\textwidth]{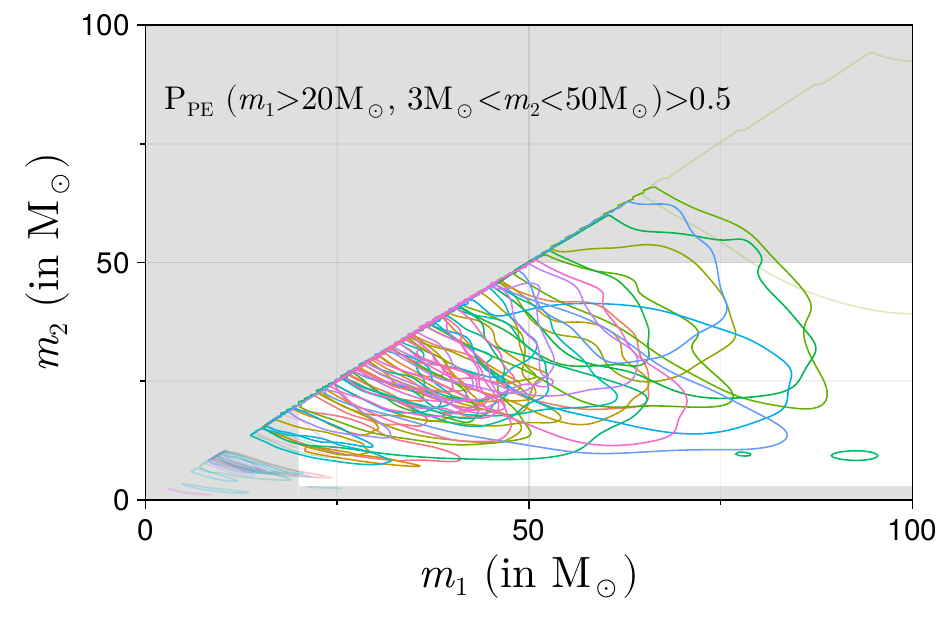}

    \includegraphics[width=0.48\textwidth]{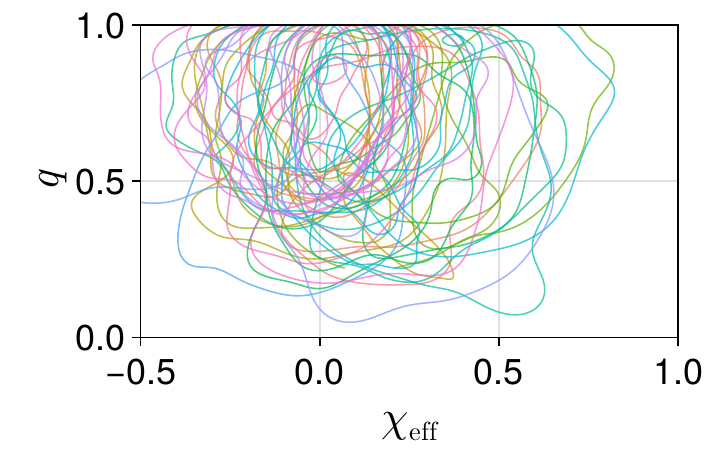} 
    \includegraphics[width=0.48\textwidth]{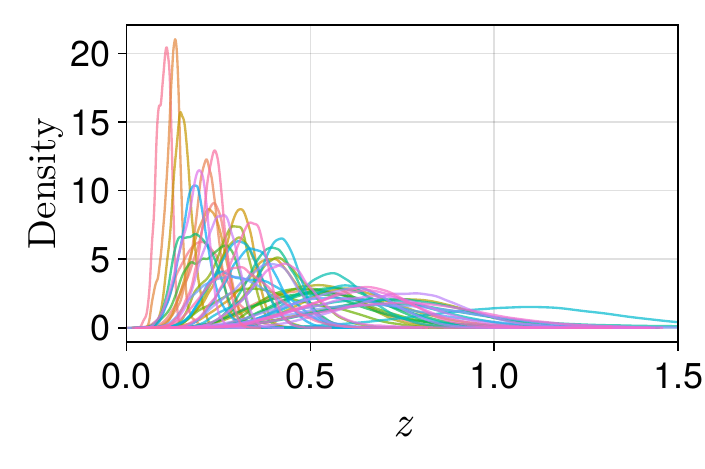}
    
\caption{$90\%$ credible contours using the default \edit{parameter estimation (PE)} prior for individual
    BBH mergers from GWTC-3. {\bf Top Panel}: Contours in the $m_1$--$m_2$ plane.  Events whose posteriors have a $50\%$
    or higher probability of $m_1 > 20 \Msun$ and $3 \Msun < m_2 < 50 \Msun$ are
    shown in full opacity; the remaining
    events are shown in $25\%$ opacity.  
    The grey region is the complement of our selected mass range.
    {\bf Bottom left panel}: The mass ratios and effective spins of the selected events.
    {\bf Bottom right Panel}: Redshifts of the selected events. See the \href{https://github.com/SoumendraRoy/35Msun_GWTC3/blob/main/m1m2_cut/scripts/Plots_Main.ipynb}{code} used to generate this plot.
    }
    \label{fig: selection}
\end{figure*}

Let $\vec{\theta} = \{\vec{\theta}^{\prime}, \vec{\theta}^{\prime\prime}\}$ denote the full set of merging BBH parameters. 
Here, $\vec{\theta}^{\prime}$ represents the source-frame masses, on which we apply selection cuts, while $\vec{\theta}^{\prime\prime}$ includes the remaining parameters, e.g.\ spins and redshift. 
Our goal is to select a specific region in mass space near the observed $35 \, \Msun$ peak.
Simultaneously, we want to ensure that our catalog is sufficiently pure of noise-contamination \citep[for an alternative approach, see][]{Roulet2020,Galaudage2020}.

We select events from the GWTC-3 event set based on the following two selection
criteria: 
\begin{enumerate}
    \item \textbf{det1:} Each trigger must have a false alarm rate (FAR) below
    % (more significant than) 
    a preassigned threshold $\mathrm{FAR}_0=1\,\mathrm{yr}^{-1}$. 
    
    \item \textbf{det2:} 
    We require that the probability of the mass parameters $\vec{\theta}^\prime$ lying within a specified region $\Omega$ exceeds a chosen threshold, $p_\mathrm{cut}$. To compute these probabilities, we use the standard \texttt{C01:nocosmo} parameter estimation (PE) samples released by the LVK Collaboration under the default prior \citep{gwtc-3pe}:
    \begin{equation}
    P_\mathrm{PE}\left( \vec{\theta}^\prime \in \Omega \right) > p_\mathrm{cut},
    \end{equation}
    with 
    \begin{multline}
        \Omega \equiv \left\{ \left( m_1, m_2 \right) \mid \right. m_1 > 20 \Msun,  \\ 
         \left. 3 \, \Msun < m_2 < 50 \, \Msun \right\}.
    \end{multline}

\end{enumerate}
    
Our default threshold is $p_\mathrm{cut} = 0.5$, but we also present analyses with $p_\mathrm{cut} = 0.1$ and $0.9$ in \ref{appen: different Mass Cut}. Additionally, we consider selection cuts based on an alternative region, $\Omega'$, defined in terms of the chirp mass, $\Mchirp$ (also in \ref{appen: different Mass Cut}). 

The conditions that define $\Omega$ ensure that at least one of the binary
components is near the $35 \Msun$ peak but exclude events that are likely to be
neutron star-black hole mergers.  
The selection is indicated in Fig.~\ref{fig: selection}. $51$ BBH merger
events\footnote{See
\href{https://github.com/SoumendraRoy/35Msun_GWTC3/blob/main/m1m2_cut/data/analyzed_gw_event_names.txt}{this
link} for the names of the analyzed GW events.} pass both selection cuts.  

The hierarchical likelihood for a population model with hyperparameters
$\vec{\lambda}$ for a catalog of events subject to the above selection
with data $\vec{d}$ is \citep{popgw2, popgw1}:

\begin{equation}\label{eq:likelihood}
    \mathcal{L}(\vec{d}|\vec{\lambda}) =  e^{-N_{\mathrm{det1,det2}}(\vec{\lambda})} \prod_{i=1}^{N_{\mathrm{obs}}} \int \dd \vec{\theta} ~ \frac{P_\mathrm{PE}(\vec{\theta}|d_i)}{\pi_{\mathrm{PE}}(\vec{\theta})} ~ \frac{\mathrm{d}N}{\mathrm{d}\vec{\theta}}(\vec{\lambda}),
\end{equation}
where $\pi_{\mathrm{PE}}(\vec{\theta})$ is the PE prior on $\vec{\theta}$.
$N_{\mathrm{det1,det2}}$ represents the expected number of BBH mergers after applying selection cuts det1 and det2:
\begin{equation}\label{eq:ndet}
    N_{\mathrm{det1,det2}} = \int \dd \vec{\theta} ~ P_{\mathrm{det1, det2}}(\vec{\theta}) ~ \frac{\mathrm{d}N}{\mathrm{d}\vec{\theta}}(\vec{\lambda}).
\end{equation}
where $P_{\mathrm{det1, det2}}(\vec{\theta})$ is the probability that an event with parameters $\vec{\theta}$ passes our selection criteria. Note that $P_{\mathrm{det1, det2}}(\vec{\theta})$ depends on $\vec{\theta}$ through \edit{the} data.

The first selection criterion is widely used in the GW literature, including the LVK population inference
\citep[e.g.][]{popgw1, gwtc-3pop}.  
% Our choice $\mathrm{FAR}_0 = 1\mathrm{yr}^{-1}$ matches 
In the second criterion, $p_\mathrm{cut}$ controls
how selective we are; setting $p_\mathrm{cut} = 0$ includes all events, while $p_\mathrm{cut} = 1$
excludes all events.  Notably, the hierarchical likelihood in
Eq.~\ref{eq:likelihood} ensures that the inference of population hyperparameters
remains unbiased for any value of $p_\mathrm{cut}$, provided that the population model
$\mathrm{d}N/\mathrm{d}\vec{\theta}(\vec{\lambda})$ adequately describes the selected
catalog of BBH mergers.

We write the $\vec{\theta}$ integrals in Eq.~\ref{eq:likelihood} as the sum over discrete PE samples drawn from the posterior $P(\vec{\theta}|d_i)$:
\begin{equation}\label{eq:likelihood1}
    \mathcal{L}(\vec{d}|\vec{\lambda}) \approx e^{-N_{\mathrm{det1,det2}}(\vec{\lambda})} \prod_{i=1}^{N_{\mathrm{obs}}} \sum_{\vec{\theta} \sim P(\vec{\theta}|d_i)} ~ \frac{\frac{\mathrm{d}N}{\mathrm{d}\vec{\theta}}(\vec{\lambda})}{\pi_{\mathrm{PE}}(\vec{\theta})} 
\end{equation}
and over detected injections drawn from a fiducial population, $p_\mathrm{draw}$
\citep{GWTC3-sensitivity-data}:
\begin{equation}\label{eq:Ndet1det2}
    N_{\mathrm{det1, det2}} \approx \frac{1}{N_\mathrm{draw}} \sum_{\vec{\theta} \sim P(\vec{\theta}|\mathrm{det1, det2, draw})} ~ \frac{\frac{\mathrm{d}N}{\mathrm{d}\vec{\theta}}(\vec{\lambda})}{p_{\mathrm{draw}}(\vec{\theta})},
\end{equation}
We ensure the convergence of both Eq.~\ref{eq:likelihood1} and \ref{eq:Ndet1det2} by guaranteeing a sufficient number of effective samples, following the approach outlined in \cite{Tiwari_2018, Pdet1-Farr, gwtc-3pop, Pdet2-essick, 10.1093/mnras/stad2968}.
Implementing our first selection criterion in these detected injections is straightforward, as the
injection set reports the measured FAR from each pipeline for each injection. 
Ideally, we would run PE on the data from each injection to get posteriors, and then apply our second detection criterion.
% To implement our second criterion, we would ideally subject perform PE on the data corresponding to each injection to obtain posteriors, and then implement det2.  
However, this process is extremely computationally costly. 
As a practical alternative, we use ``mock PE'', which provides an approximate but efficient way to estimate the actual PE posteriors for each injection \citep{Fishbach_2020,
FairhustSimplePE, Farah_2023, essick2023dagnabbitensuringconsistencynoise,
Roy_2025}. For details, see \ref{appen: mock PE}.

\subsection[Population Model for the 35Msun Peak]{Population Model for the $35\Msun$ Peak}\label{subsec:popmodel_35}

Our population model for the sources near the $35 \, \Msun$ peak consists of (i) a
``remnant'' mass function with bumpy structure drawing heavily upon
\citet{Golomb2024} that is applied to both components of the binary; (ii) a pairing
function \citep{Picky_Partners_Fishbach_2020} that allows for correlation
between the primary and secondary masses favoring or disfavoring high total
mass; (iii) a Gaussian shape for the distribution of the effective spin
\citep{Callister2021,Miller2020} with a mean and standard deviation that depend
on the mass ratio; and (iv) a redshift evolution of the merger rate that follows a
parameterized version of the \citet{MadauDickinson2014} star formation rate.  We
write 
\begin{multline}
    m_1 m_2 \frac{\dd N}{\dd m_1 \dd m_2 \dd \chi_\mathrm{eff} \dd V \dd t} = \\ R_0 f\left( m_1 \right) f\left( m_2 \right) g\left( q \right) h\left( \chi_\mathrm{eff}, q \right) r\left( z \right),
\end{multline}
with $R_0$ a merger rate density, $f$ the ``remnant'' mass function, $g$ the
pairing function, $h$ the $q$-dependent $\chi_\mathrm{eff}$ function, and $r(z)$
the evolution of the merger rate with redshift, with all frame-dependent
quantities evaluated in the source frame.  We adopt (in code) a normalization
convention where $f$, $g$, $h$, and $r$ are unitless, so that the $R_0$ parameter
has units of $\mathrm{Gpc}^{-3} \, \mathrm{yr}^{-1}$. We normalize $f$, $g$,
$h$, and $r$ so that they are 1 at a reference mass, mass ratio,
$\chi_\mathrm{eff}$, and $z$, and $R_0$ measures the merger rate density at
those reference points.  In this work, we either show properly normalized rate
densities, or \edit{explicitly} state the normalization convention.

\begin{figure}
    \centering
    \includegraphics[width=0.495\textwidth]{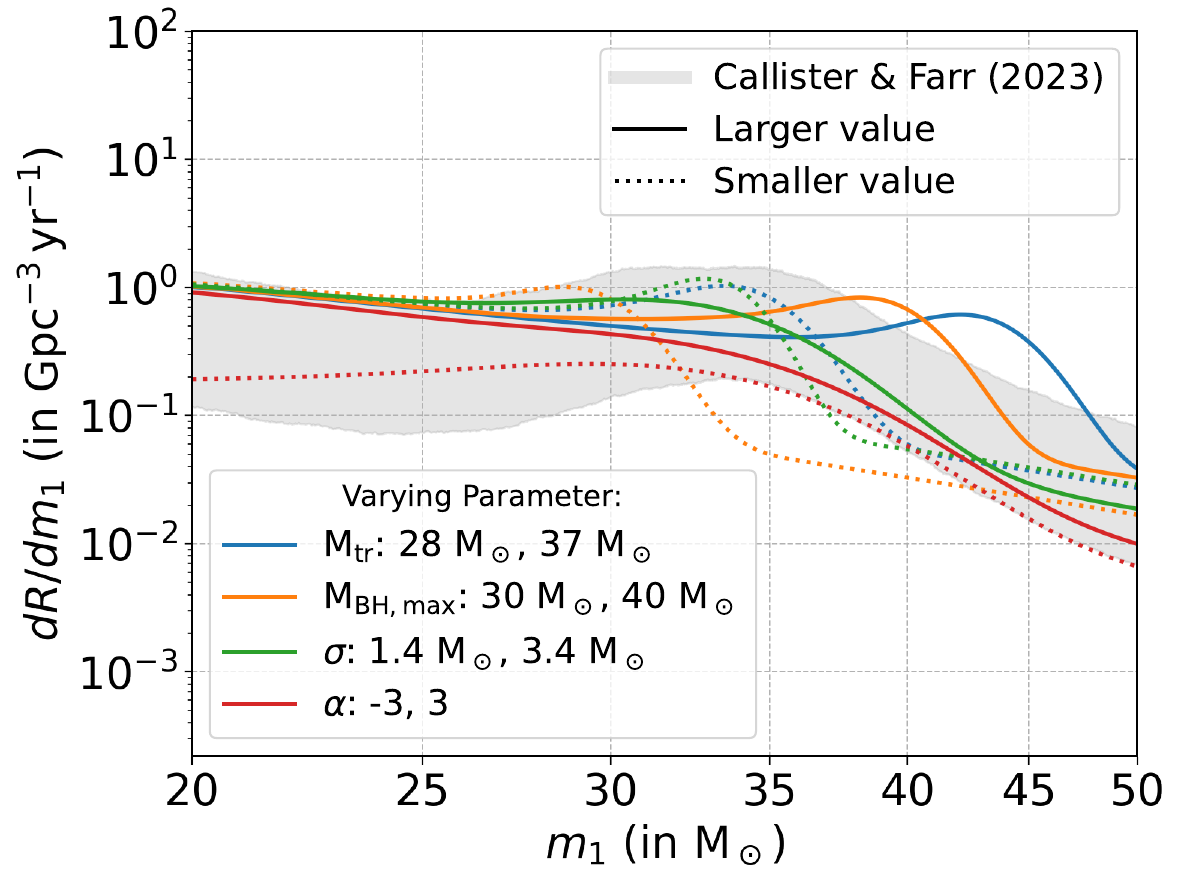}
    \caption{Behavior of our mass model (colored lines), compared to the non-parametric fit of GWTC-3 (in gray) from \cite{autoreg1}.
    In each case, we vary one parameter while keeping the others fixed at reference values: $\{\mathrm{M}_{\mathrm{tr}}=28\,\mathrm{M}_{\odot}, ~\mathrm{M}_{\mathrm{BH,max}}=34\,\mathrm{M}_{\odot}, ~\sigma=1.75\,\mathrm{M}_{\odot}, ~\alpha=2\}$. The different parameter choices for our model can reproduce the shape of the features observed in the non-parametric model. See the \href{https://github.com/SoumendraRoy/35Msun_GWTC3/blob/main/paper_plots/codes/compare_ar1.ipynb}{code} used to generate this plot.
    }
    \label{fig: structure}
\end{figure}

\begin{figure*}
    \centering

    \includegraphics[width=\textwidth]{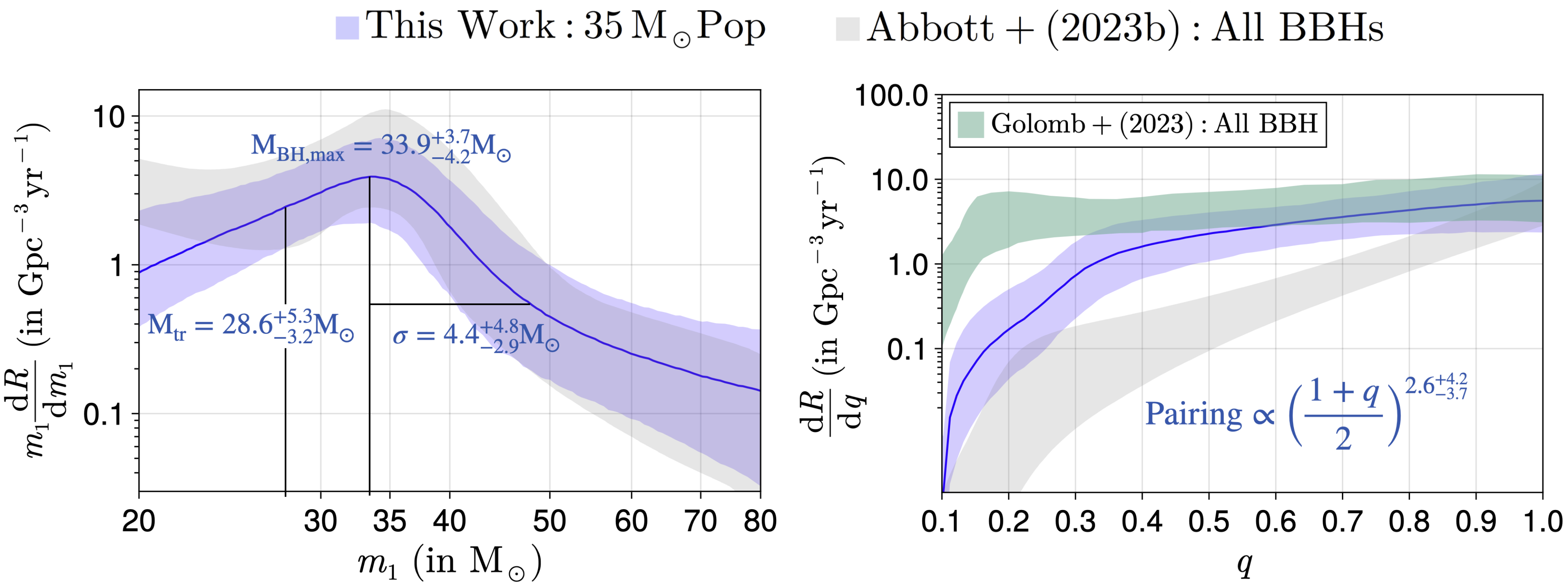}

    \includegraphics[width=\textwidth]{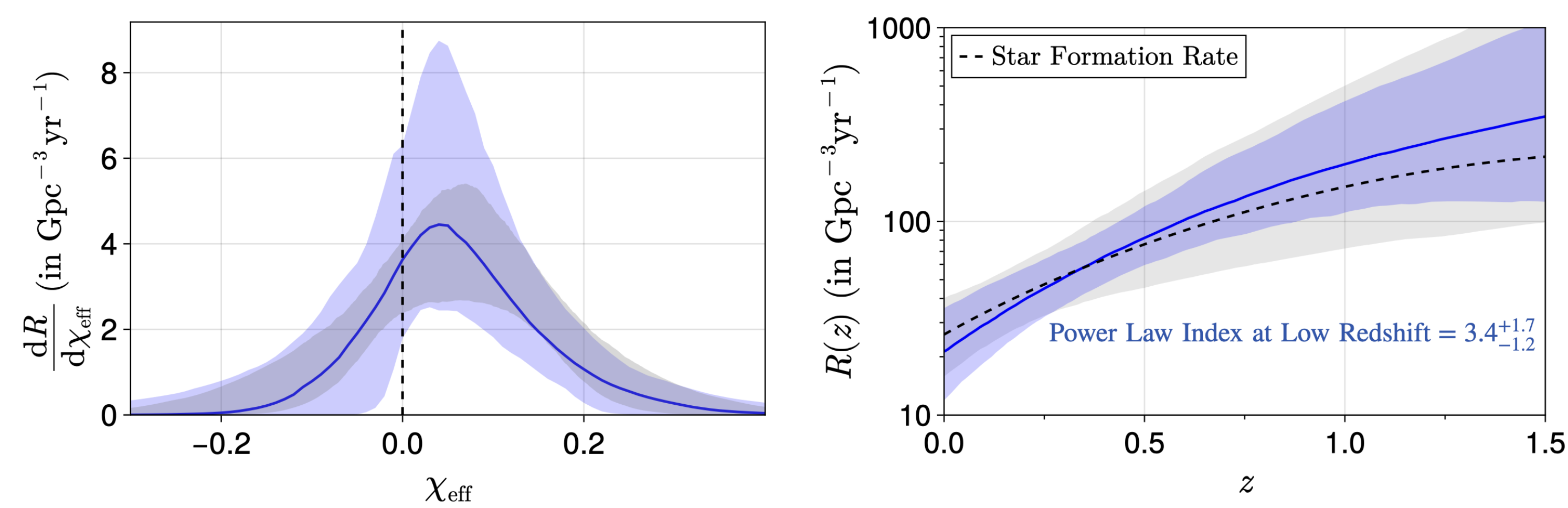}

    \caption{The primary mass, mass ratio, effective spin and redshift distributions for the $35\Msun$ peak population (blue) compared to the model from \citet{gwtc-3pop} (grey).  The $m_1$, $q$ and $\chieff$ distributions are evaluated at $z=0$. We normalize all $q$ distributions at $q = 1$ and all redshift ($z$) distributions at the value of $z$, where our measurement has the smallest error bar. 
    Panels are annotated schematically with inferences for parameters controlling the shape of the corresponding curves.
    We compare our results to the full distribution of BBHs from the Power Law + Peak model \citep[in grey][]{gwtc-3pop}.
    We find the BBH merger rate follows \edit{a} flat or shallow rise with $m_1$, increasing by a factor of 3 from $20\Msun$ up to a peak at $\MBHmax\sim 34\,\Msun$, followed by a steep drop decreasing by an order of magnitude between the peak and $50\Msun$ (top left panel).
    The $q$ distribution shows weak preference towards equal-mass mergers compared to the full population in \cite{Golomb2024}. A different pairing in \cite{gwtc-3pop} shows \edit{a} steeper $q$ population than \edit{ours}.
    The $\chieff$ distribution is skewed positively (a median of $0$ is ruled out at $\sim 90\%$ confidence). 
    The merger rate for the $35\Msun$ peak population increases with $z$, at a rate that is consistent with the low-redshift behavior of the star formation rate (the dashed line). See the \href{https://github.com/SoumendraRoy/35Msun_GWTC3/blob/main/m1m2_cut/scripts/Plots_Main.ipynb}{code} used to generate this plot.}
    \label{fig: main results}
\end{figure*}

The ``remnant'' mass function follows closely the ``bumpy'' model from
\citet{Golomb2024}, which we reproduce here for clarity.  It is composed
of two pieces, and has six parameters whose meanings are described below:
\begin{multline}
    f\left(m \mid \alpha, \Mtr, \MBHmax, \sigma, \alpha_h \right) = \\ f_b\left(m \mid \alpha, \Mtr, \MBHmax, \sigma \right) + \lambda_h f_h\left(m \mid \alpha_h \right).
\end{multline}
The authors imagine that there is a power-law
``initial'' mass function $p\left( m_i \right) \propto m_i^{-\alpha}$ that is
passed through a stochastic process relating initial to remnant masses.  
We assume that this process produces a Gaussian distribution of remnant masses for each initial mass, with 
\begin{equation}
p\left( m \mid m_i \right) = N\left[ \mu_m\left( m_i \mid \Mtr, \MBHmax \right), \sigma \right]\left( m \right).
\end{equation}
The mean of the Gaussian distribution of final masses is given as a function of
initial mass by 
\begin{multline}
\mu_m\left( m \mid \Mtr, \MBHmax \right) = \\ \begin{cases}
    m & m < \Mtr \\
    \MBHmax - \frac{\left( m - \left( 2 \MBHmax - \Mtr \right) \right)^2}{4 \left( \MBHmax - \Mtr \right)} & m \geq \Mtr
\end{cases};
\end{multline}
the $\sigma$ parameter controls the scatter. For masses below the transition
mass, $m < \Mtr$, the mean is linear in $m$, transitioning smoothly to a
quadratic that peaks with remnant mass $\MBHmax$ for masses $m > \Mtr$.  
The turnover in $\mu_m$ leads to a pile-up of remnants near $\MBHmax$. The sharpness of this peak, as well as \edit{is} the decay of the remnant mass function beyond it, are governed by the parameter $\sigma$. In the limit $\sigma \to 0$, the model reproduces the infinitely sharp peak described by \citet{Baxter2021}. At finite $\sigma$, the mass distribution follows a power-law for $m \lesssim \Mtr$. Between $\Mtr$ and $\MBHmax$, there is a pile-up of mergers, with the distribution peaking near $m \simeq \MBHmax$. Beyond $\MBHmax$, the distribution falls off as a Gaussian with a characteristic width set by $\sigma$. See \edit{Figs}~\ref{fig: structure}
and \ref{fig: main results}.

Because our mass cut permits \edit{arbitrarily} high primary masses, we
introduce a ``high-mass'' remnant mass function, $f_h$, which is a truncated
power law that begins at $\MBHmax$ and behaves like $m^{-\alpha_h}$ for $m
\gtrsim \MBHmax$.  The \edit{parameter} $\lambda_h$ controls the relative contribution
of this component to the remnant mass function.  

Fig.~\ref{fig: structure} shows our bumpy model across several variations of
its key shape parameters. Comparing our model to the highly flexible model from
\cite{CallisterFarr2024} demonstrates that our model is sufficiently flexible to
capture the relevant structure around the $35\Msun$ peak.

\edit{Following} \citet{Picky_Partners_Fishbach_2020}, we apply the remnant mass
function $f$ with all its structure to both $m_1$ and $m_2$; correlations
between the masses are introduced by a ``pairing function.''  Here we choose a
pairing function that is a power law in the total mass of the binary, with one
parameter:
\begin{equation}\label{eq:pairing}
    g\left(q \mid \beta \right) = \left(\frac{1+q}{2}\right)^{\beta} \edit{.}
\end{equation}
Unlike the power law in mass ratio pairing used in \citet{gwtc-3pop}, our
formulation allows for a correlation between $m_1$ and $m_2$ (i.e.\ for $\beta
\neq 0$ the mass function cannot be written as a product of a function of $m_1$
and another function of $m_2$) and imposes structure on both mass components \citep[cf.][]{farah2024kindcomparingbigsmall}. We
have also considered a Gaussian-shaped pairing function, but the results of that
analysis are essentially identical to the one reported here, so we do not
discuss it further.

Following \citet{Callister2021}, we want to allow for the possibility of
effective spins correlated with mass ratio, so we introduce an effective spin
function, $h$, with four parameters:
\begin{multline}
    h\left( \chi_\mathrm{eff}, q \mid \mu_0, \mu_1, \sigma_0, \sigma_1 \right) = \\  \mathcal{N}[\chieff | \mu_\mathrm{eff}\left( q \mid \mu_0, \mu_1 \right), \sigma_\mathrm{eff}\left( q \mid \sigma_0, \sigma_1 \right)],
\end{multline}
where $\mathcal{N}$ is the normalized probability density function of \edit{a} normal distribution with mean,
\begin{equation}
    \mu_\mathrm{eff} \left( q \mid \mu_0, \mu_1 \right) = \mu_0 \left( 1 - q \right) + \mu_1 q
\end{equation}
and width,
\begin{equation}
    \sigma_\mathrm{eff} \left( q \mid \sigma_0, \sigma_1 \right) = \sigma_0 \left( 1 -q \right) + \sigma_1 q
\end{equation}
$\mu_\mathrm{eff}$ and $\sigma_\mathrm{eff}$ are linear functions of $q$ parameterized by their values at $q=0$ and $q=1$.  With
this parameterization, the derivative of the mean and standard deviation with
$q$ are given by 
\begin{equation}
    \label{eq:alpha-eff-definition}
    \frac{\dd \mu_\mathrm{eff}}{\dd q} \equiv \alpha_{\chi_\mathrm{eff}} = \mu_1 - \mu_0
\end{equation}
and
\begin{equation}
    \label{eq:beta-eff-definition}
    \frac{\dd \sigma_\mathrm{eff}}{\dd q} \equiv \beta_{\chi_\mathrm{eff}} = \sigma_1 - \sigma_0.
\end{equation}
Here non-zero values $\alpha_{\chieff}$ and $\beta_{\chieff}$ show the $q-\chieff$ correlation.

We assume that the other dimensions of the spin parameter space
follow a uniform distribution in magnitude and an isotropic distribution of
angles, conditional on $\chi_\mathrm{eff}$ \citep{Callister2021Thesaurus}.  This
parameterization does not assume any more complex or highly structured
correlation between $q$ and $\chieff$  \citep[see, the discussions
in][]{Callister2021, spin_ev, Adamcewicz_2022, Adamcewicz_2023, gwtc-3pop,
PhysRevD.109.103006}. 

We allow the merger rate to evolve in redshift via a redshift function, $r$,
with three parameters:
\begin{equation}
    r\left(z \mid \lambda, z_p, \kappa \right) \propto \frac{\left( 1 + z \right)^\lambda}{1 + \left( \frac{1 + z}{1 + z_p} \right)^\kappa}.
    \label{eq: srf}
\end{equation}
For the specific choice $\lambda = 2.7$, $z_p = 1.9$, and $\kappa = 5.6$, this
follows the shape of the star formation rate \edit{(SFR)} inferred by
\citet{MadauDickinson2014}.  The bottom-right panel of Fig.~\ref{fig: selection}
shows that all BBH mergers occur at redshifts well below $2$, a regime where the
star formation rate is known to increase with redshift 
(\citealt{MadauDickinson2014, Vangioni_2015, Ghirlanda_2016}, though cf. \citealt{2023ApJ...957L..31F}).
We note that BBH merger rates may not strictly follow the shape of the global SFR evolution (Eq.~\ref{eq: srf}), as the redshift evolution of the merger rate can be dominated by formation-channel-dependent delay times \citep[see e.g.,][]{2021ApJ...914L..30F, vanson2022_rate_redshift, boesky2024binaryblackholemerger}. %2023ApJ...957L..31F,
At the same time, BBH mergers may trace low-metallicity star formation \citep[e.g.,][and references therein]{2025ApJ...979..209V} but, the shape of the low-metallicity SFR distribution remains highly uncertain \citep[][]{2024AnP...53600170C}. This thus has a significant impact on BBH merger rate predictions \citep{2017MNRAS.472.2422M,2021MNRAS.502.4877S,2022MNRAS.516.5737B,2025A&A...698A.144S}.

% It is important to note that the star formation–motivated form of $r(z)$ may not be universally valid, as different regions of the BBH mass spectrum may exhibit distinct delay time distribution and hence, redshift evolutions \citep[see][]{redshift_ev3,CallisterFarr2024,boesky2024binaryblackholemerger, boesky2024investigatingcosmologicalratecompact}.  
% \todo{Lieke wants to look at preceding paragraph and add more citations from other before we submit to the arxiv}

%%%%%%%%%%%%%%%%%%%%%%%%%%%%%%%%%%%%%%%%%%%%%%%%%%%%%%%%%%%%%%%%%%%%%%%%%%%%%%%%%%%%%%%%%%%%%%%%%%%%%%
\section{Results}\label{sec:Result}
All error bars, including the uncertainty bands in the plots, show $90\%$ credible intervals\footnote{For model hyperpriors, see the \href{https://github.com/SoumendraRoy/35Msun_GWTC3/blob/3cf203163ee3c2a96fd9ea8a1587c3283b8a96cc/m1m2\_cut/src/model.jl\#L329-L347}{code snippet}. For the full contour plot, see this \href{https://github.com/SoumendraRoy/35Msun_GWTC3/blob/main/m1m2_cut/figures/corner_compare_selection.pdf}{link}. For the median values with $90\%$ uncertainties in all hyperparameters, see `Print Median, 5\% and 95\% Values' block of this \href{https://github.com/SoumendraRoy/35Msun_GWTC3/blob/main/m1m2_cut/scripts/Plots_Main.ipynb}{Jupyter Notebook}.}.

%%%%%%%%%%%%%%%%%%%%%%%%%%%%%%%%%%%%%%%%%%%%%%%%%%%%%%%%%%%%%%%%%%%%%%%%%%%%%%%%%%%%%%%%%%%%%%%%%%%%%%
%%%%%%%%%%%%%%%%%%%%%%%%%%%%%%%%%%%%%%%%%%%%%%%%%%%%%%%%%%%%%%%%%%%%%%%%%%%%%%%%%%%%%%%%%%%%%%%%%%%%%%

%% MAIN RESULTS FIGURE

%%%%%%%%%%%%%%%%%%%%%%%%%%%%%%%%%%%%%%%%%%%%%%%%%%%%%%%%%%%%%%%%%%%%%%%%%%%%%%%%%%%%%%%%%%%%%%%%%%%%%%
%%%%%%%%%%%%%%%%%%%%%%%%%%%%%%%%%%%%%%%%%%%%%%%%%%%%%%%%%%%%%%%%%%%%%%%%%%%%%%%%%%%%%%%%%%%%%%%%%%%%%%

\subsection{Primary Mass:}\label{subsec: Primary Mass Result}
In the top left panel of Fig.~\ref{fig: main results}, we find that the distribution of the primary mass, $m_1 \mathrm{d}N/\mathrm{d} m_1$, is flat or rising with $m_1$ at low masses. The distribution starts to deviate from a power-law around $28.6^{+5.3}_{-3.2} \Msun$, reaches a peak at $33.9^{+3.7}_{-4.2}\Msun$, and has a narrow width of $4.4^{+4.8}_{-2.9}\Msun$, before dropping by about an order of magnitude by $50\Msun$. 

At approximately $20\Msun$, we observe a significant deviation from the `Power
Law + Peak' fit by \cite{gwtc-3pop} for the full primary mass distribution of
all BBH merger events. This deviation is expected, as we did not model the
low-mass end of the mass function, and so do not need to have a model component
that rises toward the global peak of the mass function at $\sim 10 \, \Msun$ as
observed in \citet{gwtc-3pop}.  However, in the vicinity of the $35 \Msun$ peak,
the `Power Law + Peak' model for all BBH mergers aligns well with our results.
This agreement suggests that modeling the region above $m_1>20 \Msun$ is
sufficient to capture the structure of the $35 \Msun$ peak reliably. We conclude
that our findings regarding the primary mass distribution near the $35 \Msun$
peak are largely consistent with those of \cite{gwtc-3pop} and
\cite{Golomb2024}.

%%%%%%%%%%%%%%%%%%%%%%%%%%
\subsection{Mass Ratio}\label{subsec:Mass-Ratio-Result}

In the top right panel of Fig.~\ref{fig: main results}, we find that the
power-law index of the pairing function is $2.6^{+4.2}_{-3.7}$. Even though the
pairing function power law index can be positive or negative by prior distribution, the marginal
mass-ratio distribution shows a clear preference for equal-mass mergers (see \cite{Li_2022,Li_2024_apj} for a similar claim).

Compared to the marginal $q$ distribution for the full BBH merger catalog
reported in \cite{gwtc-3pop}, our $q$ distribution appears relatively more flat.
This trend is consistent with the findings of \cite{Golomb2024} (green line in
top right panel of \ref{fig: main results}). The form of the pairing function in
\citet{gwtc-3pop} is different than ours and that of \citet{Golomb2024}.  Our
analysis uses a joint mass distribution of the form $P(m_1, m_2) = f(m_1) f(m_2)
(\frac{1+q}{2})^{\beta}$, which introduces correlations between $m_1$ and $m_2$.
In contrast, \cite{gwtc-3pop} uses a simpler pairing model that assumes no
correlation between $m_1$ and $m_2$, adopting $P(m_1, m_2) = \tilde{f}(m_1)
\tilde{g}\left( m_2 \right)$. 
When comparing to \cite{Golomb2024} (which uses the same mass function and pairing), we find that the $35\Msun$ peak populations \edit{prefer} equal mass mergers more strongly than the full BBH merger catalog. Note that our broad $m_2$ range ($3–50 \Msun$) is comparable to \cite{Golomb2024}, except that our $m_2$ cut excludes GW190521 \citep{GW190521_Abbott_2020} (which anyway prefers equal mass). On the low-mass end, our stricter $m_1 > 20 \Msun$ cut allows for small $m_2$ and large $m_1$ events, so the dip at low $q$ comparable to \cite{Golomb2024} is not due to our mass cuts\footnote{For an equivalent population distribution of secondary mass, marginalized over $m_1$, and evaluated at $\chieff=0$ and $z=0$, see \href{https://github.com/SoumendraRoy/35Msun_GWTC3/blob/main/m1m2_cut/scripts/Plots_More.ipynb}{this notebook}.}.

We also find that the marginal mass-ratio distribution is sensitive to the
specific BBH merger events included in the analysis. Applying a stricter
selection cut (see \ref{appen: different Mass Cut}) removes several low-mass
systems with extreme mass ratios (in particular GW190412 \citep{Abbott_2020_GW190412}) which leads to a stronger
preference for equal-mass mergers. 

%%%%%%%%%%%%%%%%%%%%%%%%%%
\subsection{Effective Spin}\label{subsec:Effective-Spin-Result}

\begin{figure*}
    \centering

    \includegraphics[width=\textwidth]{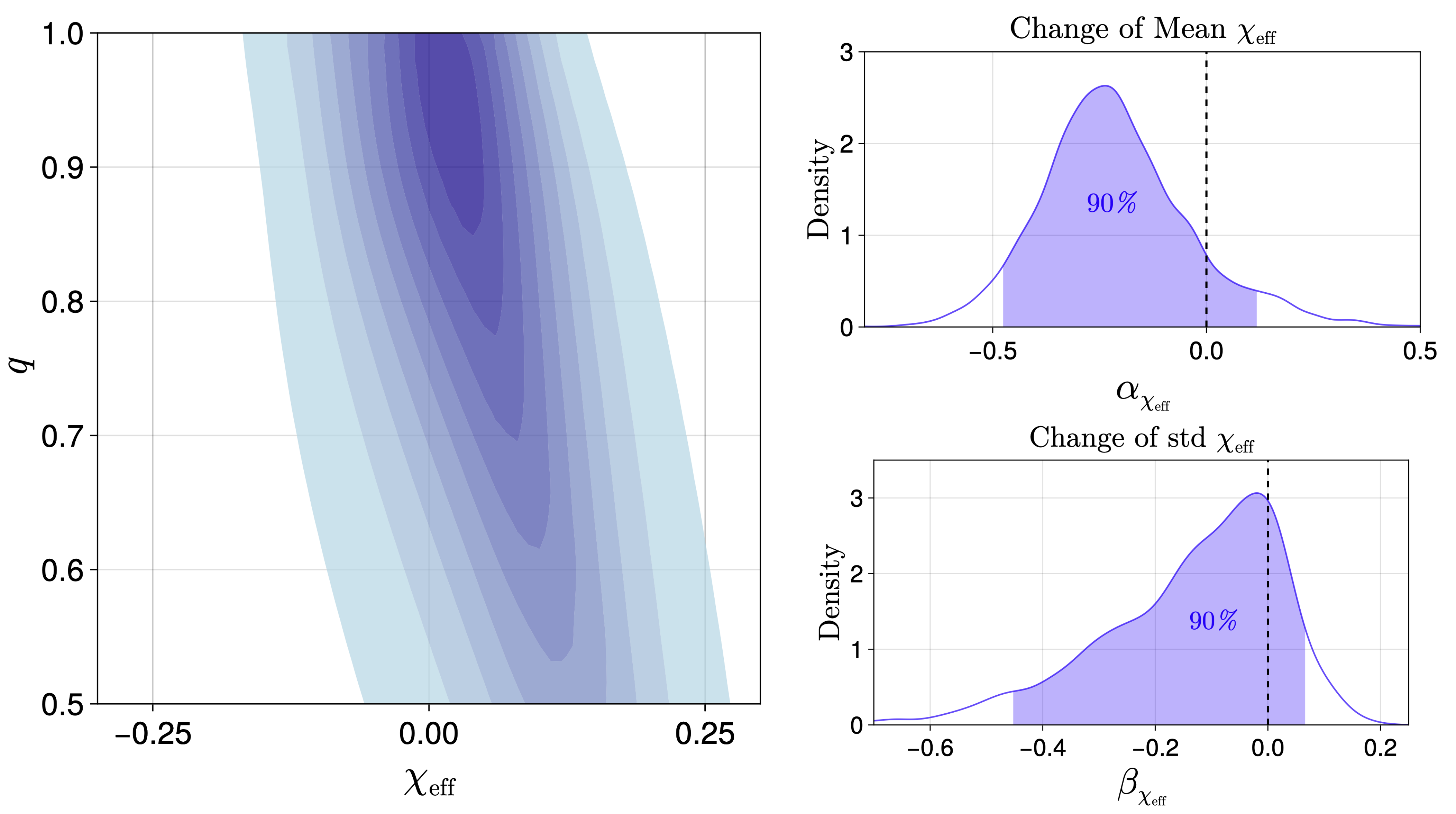}
    
    \caption{{\bf Left panel:} Joint population distribution of mass ratio and
    effective spin for the $35\Msun$ peak population, evaluated at $m_1=35\Msun$
    and $z=0$. {\bf Right panel:} Marginal distributions for $\alpha_{\chieff}$
    (variation of the mean of $\chieff$) Eq.~\ref{eq:alpha-eff-definition}, and
    $\beta_{\chieff}$ (variation of the width of $\chieff$),
    Eq.~\ref{eq:beta-eff-definition}.
    Although most of the posterior support lies at $\alpha_{\chieff}<0$, our
    analysis remains consistent with no evolution, $\alpha_{\chieff},
    \beta_{\chieff} = 0$, within the $90\%$ credible interval. We find significant posterior support at $q=0.5$ in reweighted PEs. See the \href{https://github.com/SoumendraRoy/35Msun_GWTC3/blob/main/m1m2_cut/scripts/Plots_Main.ipynb}{code} used to generate this plot.}
    \label{fig: q-chieff pop}
\end{figure*}

In the bottom left panel of Fig.~\ref{fig: main results}, we find that the marginal $\chieff$ distribution prefers positive values. The median value of $\chieff$ is larger than zero at around $90\%$ credible interval. 
We also find the width of marginal $\chieff$ distribution is around $\sim 0.1$. Both median and width are consistent with the full BBH merger population \citep{gwtc-3pop}  (for comparison with other existing results, see \ref{appen: compare chieff}).
Unlike the mass-ratio analysis, this result is robust against all variations of the mass cut we explored in this paper (see Fig.~\ref{fig: diff sel}). 

%%%%%%%%%%%%%%%%%%%%%%%%%%
\subsection{Mass Ratio - Effective Spin Correlation}\label{subsec:q-chieff Result}

\citet{Callister2021} and \cite{gwtc-3pop} observe an anti-correlation between
mass ratio and mean effective spin in the full BBH merger population.  In
Fig.~\ref{fig: q-chieff pop}, we show the joint distribution of mass ratio and
effective spin for the $35\Msun$ peak population. We observe that both the mean
and standard deviation of the effective spin may decrease with the increasing
mass ratio (more unequal mergers have a larger mean and wider standard deviation
in $\chi_\mathrm{eff}$), but our inference is consistent with no change within
the $90\%$ credible interval.  
The results \edit{of} how the width of the $\chieff$ distribution changes with $q$ are also sensitive to which events are included in our analysis
(\ref{appen: different Mass Cut}).  Our inference for the change in mean and
standard deviation of the $\chieff$ distribution ($\alpha_{\chieff}$ and $\beta_{\chieff}$ respectively) is consistent with all BBH
mergers in GWTC-3.  However, likely due to \edit{the} smaller number of events with good
$q-\chieff$ constraints in our subset, we cannot distinguish between an
increasing, decreasing, or flat relationship between $\chieff$ and $q$ at high confidence.

These results suggest that the catalog-wide anti-correlation between $q$ and $\chieff$ \citep{gwtc-3pop} may originate from low-mass systems ($m_1<20\Msun$) or from mixing distinct subpopulations--for example, if low-mass events preferentially involve unequal-mass mergers with higher $\chieff$, while high-mass events favor equal-mass mergers with lower $\chieff$. We leave a more detailed investigation of this effect to future work.

%%%%%%%%%%%%%%%%%%%%%%%%%%
\subsection{Redshift}\label{subsec:Redshift Result}

In the bottom right panel of Fig.~\ref{fig: main results}, we show the inferred redshift evolution of the systems in the $35\Msun$ peak. We find that the BBH merger rate associated with the $35\Msun$ peak population increases with $z$, characterized by a low-$z$ power-law index of $\lambda=3.4^{+1.7}_{-1.2}$. 
This firmly rules out scenarios in which the BBH merger rate is constant or decreases with redshift up to $z=1.5$, with a significance of approximately $4\sigma$. 
Our result is consistent with the evolution of the cosmic star formation rate \citep[\edit{SFR,}][]{MadauDickinson2014, Madau_2017}, as well as with the trend observed for all BBH mergers in the GWTC-3 \citep{gwtc-3pop}.
Interestingly, our findings are more constraining on the rate evolution with redshift than that of the all BBH merger population.  It is possible this arises because the full population is a mixture of different redshift evolutions of low- and high-mass systems \citep{autoreg1, Lalleman_2025}, and not well fit by a smooth power law.  Thus, considering only the $35\Msun$ population may have more constraining power.

%%%%%%%%%%%%%%%%%%%%%%%%%%%%%%%%%%%%%%%%%%%%%%%%%%%%%%%%%%%%%%%%%%%%%%%%%%%%%%%%%%%%%%%%%%%%%%%%%%%%%%
\section{Astrophysical origin of the $35\Msun$ feature }\label{sec:astrointerp} 
\begin{figure*}
    \includegraphics[width=\linewidth]{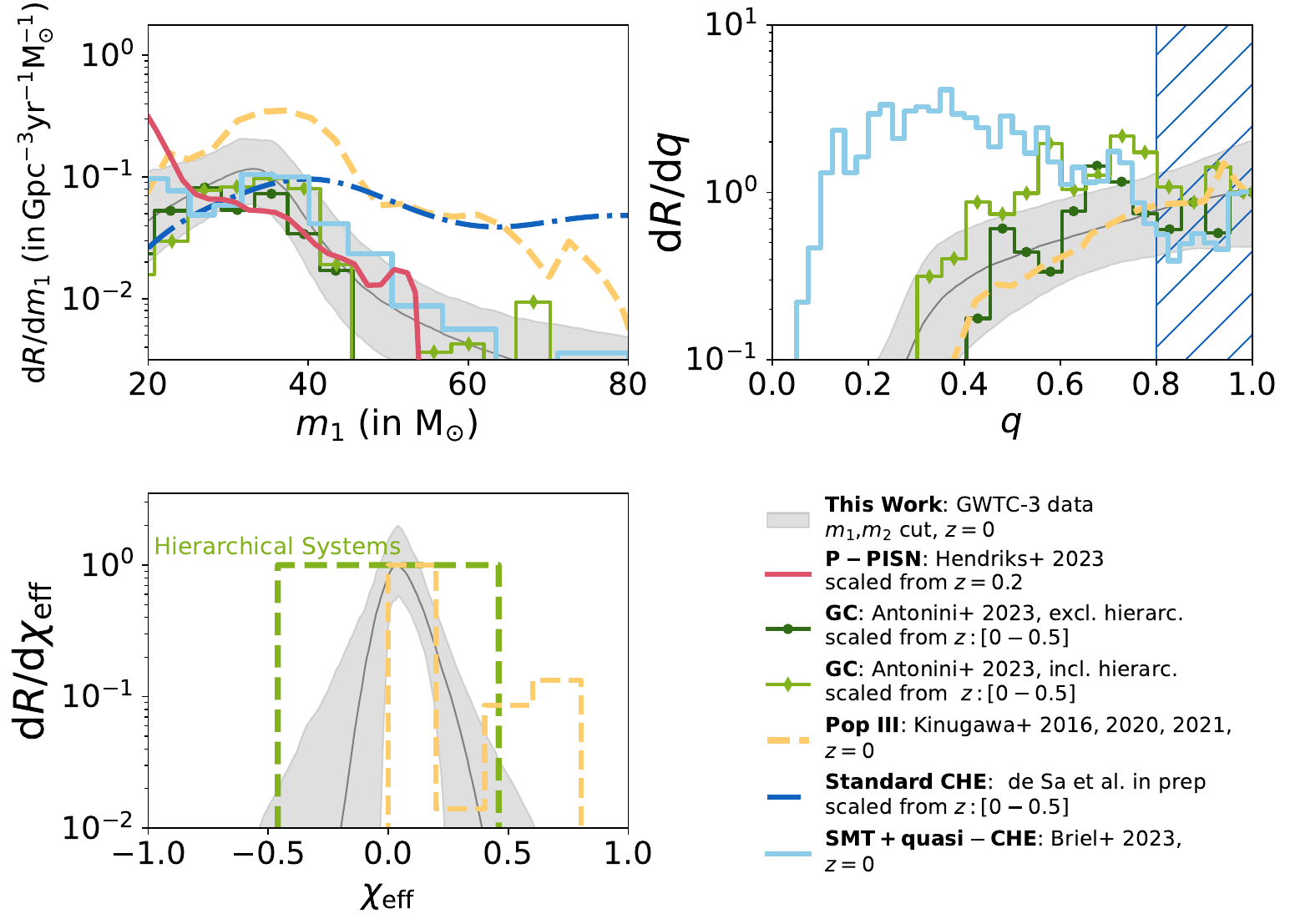} 
    \caption{
    Comparison between our results and predictions from formation channels proposed in the literature for the $35\,\Msun$ peak population.
    In the top-left panel the primary mass distributions from the literature are rescaled to redshift zero using the merger rate evolution model \edit{$R(z) = R_0(1+z)^{\lambda}$} with $\lambda = 3.4$ \citep{MadauDickinson2014}. %Madau_2017
    The mass ratio distributions are normalized to a median of 1 at $q = 1$ (top right). The hatched blue region shows $q>0.8$ for the CHE channel \citep[e.g.,][]{Hastings_2020}.
    The effective spin distributions are normalized to the peak value (bottom left).
    The dashed green line shows the prediction for hierarchical mergers as discussed in Section \ref{ss: hierarchical mergers}.
    While several formation channels reproduce the location of the observed $\sim35\Msun$ peak in $m_1$, none of them simultaneously provide a satisfactory match to the observed peak population in $m_1$, $q$, and $\chieff$. See the \href{https://github.com/SoumendraRoy/35Msun_GWTC3/blob/main/codes/scripts/Compare_formation_channels.ipynb}{code} used to generate this plot.
    } 
    \label{fig:compare form channels} % Add a label for referencing
\end{figure*}

Next, we compare our findings for $m_1$, $q$, and $\chieff$ to proposed formation scenarios from the literature in Fig.~\ref{fig:compare form channels}. 
Some formation channels also provide predictions for the evolution of the rate with redshift \citep[e.g.,][de Sa et al., in preparation]{dynamic1,2021RNAAS...5...19R}. 
However\edit{,} we have chosen not to show these in Fig.~\ref{fig:compare form channels} since the models that do provide predictions are all consistent with our inferences for the low-$z$ evolution of the merger rate. 
%

%%%%%%%%%%%%%%%%%%%%%%%%%%%%%%%%

%%%%%%%%%%%%%%%%%%%%%%%%%%%%%%%%
\subsection{Pair-instability pulsations \label{ss: PPISN}}

One of the main prevailing theories is that the $35\Msun$ feature in the black hole mass distribution can be attributed to a pile-up from P-PISN (see Section \ref{sec:Intro}). However, this interpretation is at odds with several theoretical predictions. 
Firstly, stellar models consistently predict the maximum BH mass (and thus the location of the P-PISN induced pile up) to lie between $45$ and $103\Msun$ \citep[][pink line in top left panel of Figure \ref{fig:compare form channels}]{Woosley_2002,Woosley2017,Marchant2019,Farmer_2019,Farmer2020,Renzo2020_conv_edge_ppisn,Farag2022,Hendriks2023}.
This location is relatively robust to most stellar evolution uncertainties, with only variations in the $^{12}$C$(\alpha,\gamma)^{16}$O reaction rate causing significant shifts in the pile-up location \citep{Farmer_2019,Farmer2020}.
While a $-3\sigma$ deviation in the $^{12}$C$(\alpha,\gamma)^{16}$O rate can shift the pile-up up to $\sim100\Msun$, \textit{increasing} the rate has a much smaller effect: even at the $+3\sigma$ value of the $^{12}$C$(\alpha,\gamma)^{16}$O rate, the location of the pile up never drops below $\sim45\Msun$ \citep{Farag2022}. 
\footnote{While some earlier studies \citep[e.g.,][]{Farmer_2019,Marchant2019} predicted the pile-up to occur near $40\Msun$, these results were based on unresolved implementations of the $^{12}$C$(\alpha,\gamma)^{16}$O reaction rate \citep{Mehta2022}. None of the recent models using updated rates predict a maximum BH mass below $45\Msun$ \citep[e.g.,][]{Shen2023}.}
Conversely, \textit{requiring} that the pile-up is at $35\Msun$ is caused by P-PISN would imply a CO rate that is unrealistic from a nuclear-physics perspective \citep{Golomb2024}.
Moreover, artificially lowering the pile-up location to match the $35\Msun$ feature increases the predicted P-PISN rate, causing a stronger tension with the scarcity of electromagnetic transients that have been robustly identified as (P-)PISNe \citep{Hendriks2023,RenzoSmith2024}.
% Also you should remember that changing the CO rate is not inconsequential, it also changes all other stars because it is a really important reaction rate for stellar evolution!

% What do the 'other properties' say
The ``other'' properties (mass ratio, effective spin, spin-orbit misalignment, and eccentricity) of BBH mergers involving a P-PISN progenitor have received considerably less attention than their masses.
Generally\edit{,} the mass ratio and eccentricity distributions depend strongly on the assumed pairing function, i.e., the formation channel. 
The most likely channels for such high-mass systems (stable mass transfer\footnote{\edit{See \citep[][]{1987ApJ...318..794H, 1997A&A...327..620S, Ge_2010, Ge_2015, Ge_2020, ge2024masstransferphysicsbinary, Pavlovskii_2015, Pavlovskii_2016}. For applications to binary formation channels, see \citep[][]{zams7, van_Son_2022, Olejak_2024, Picco_2024}.}}, chemically homogeneous evolution and dynamical formation channels), 
are described below. 
\citet{Marchant2019} further speculates that pair-pulsations could enhance the eccentricities, \citep[similar to a Blaauw kick][]{Blaauw1961}, but these eccentricities would not survive to be detectable in current GW data. 
They further note that rapid mass loss during the pulses could reduce the final BH spin by 30--50\%. 
Such low spins could be consistent with the low effective spins that we find for the $35\Msun$ feature.
If P-PISN ejections are asymmetric, stronger pulses could also induce  spin-orbit misalignment.
However, to date\edit{,} there are no multidimensional P-PISN simulations that can test the asymmetry of P-PISN ejections.
%%%

Given the aforementioned considerations, we conclude that the $35\Msun$ feature is unlikely to be caused by P-PISN in the ``traditional'' sense, i.e., a pile-up in the mass distribution of BHs formed from P-PISN progenitors.
We will refer to this as the ``standard P-PISN'' scenario ({\bf 1A} in Fig.~\ref{fig:astro table}). \\

% Variations on the PPISN story
Several variations on the standard P-PISN model have been proposed. % as an explanation for the $35\Msun$ feature. 
Firstly \citet{CroonSakstein2023} describe a ``shoulder'' at the high-mass end of the initial-to-remnant mass function (we will refer to this as the ``Shoulder P-PISN'' scenario, or {\bf 1B} in Fig.~\ref{fig:astro table}).
They find this feature can produce a peak around $35\Msun$ when a $+3\sigma$ variation in the $^{12}$C$(\alpha,\gamma)^{16}$O reaction rate is assumed. 
However, because the shoulder originates from more massive progenitor stars, it is predicted to yield a substantially lower merger rate than the primary P-PISN feature (which appears near $40\Msun$ in their $+3\sigma$ model). The absence of a secondary peak at higher BH masses therefore poses a challenge to this explanation.
Second, as noted by \cite{vanson2022_rate_redshift} \edit{and} \cite{Hendriks2023}, the transition between CCSNe and P-PISNe could result in a discontinuous remnant mass function if significant mass loss occurs during the first P-PISN pulse. 
Such a degeneracy would, in turn, lead to a pile-up near $35\Msun$. 
We refer to this as the ``P-PISN onset degen.'', or {\bf 1C} in Fig.~\ref{fig:astro table}.
However, pulses occurring near the boundary between CCSNe and P-PISNe are expected to be relatively low energy, and are thus unlikely to eject much mass \citep[cf.][]{Renzo2020_conv_edge_ppisn}.
Considerable uncertainties remain in this modeling regime, particularly because pulsations have not yet been modeled self-consistently in three dimensions.

% Similarly, \cite{Winch2025} argue that the maximum mass right \textit{before} the onset of pair-pulsations lies at a CO core mass of $M_{\rm CO,crit} = 36.3 \Msun$ which they remark is similar to the feature observed in GW. They argue this mass is reached in particular for rapidly rotating stars that evolve homogeneously, since they argue these homogeneous stars loose more mass due to added (mechanical) mass loss. However, they neglect orbital widening effects that would be induced by the mass loss, which most likely would prevent a GW merger.  

% III) Sakstein for beyond standard model stuff ?

%%%%%%%%%%%%%%%%%%%%%%%%%%%%%%%%
\subsection{Chemically homogeneously evolving stars}
% How this would create a $35\Msun$ peak
Chemically homogeneous evolution (CHE) naturally favors the formation of more massive BHs \citep[][]{deMink_2009, deMink2010}.
In this scenario, efficient rotational mixing keeps the progenitor stars compact, preventing them from filling their Roche lobe. 
CHE is especially relevant for massive stars ($\gtrsim20\Msun$), where increased radiation pressure enhances internal circulation, making rotational mixing more effective \citep[e.g.,][]{Hastings_2020}.
Of particular interest for GW progenitors is the case of close, tidally locked binaries \citep{MandeldeMink2016, Marchant2016}. 
Under the right conditions, CHE evolution could thus lead to a feature around $35\Msun$ \citep[as suggested by e.g.,][]{Sharpe_2024}.
de Sa et al. (in preparation) explore this scenario in detail and find that CHE can indeed produce a feature in the black hole mass distribution around $35 \Msun$ at $z=0$ (see dark blue dash-dotted line in Fig.~\ref{fig:compare form channels}, and the {\bf2.A} CHE-standard row in Fig.~\ref{fig:astro table}).
However, although this channel matches the ``rise to the peak'', it does not naturally predict a steep decay at higher masses, as observed in our inference.
% What do the 'other properties' say
Moreover, de Sa et al. (in preparation) note that the final core spins of these systems are expected to be high. 
This does not fit with the low effective spins observed in our results.
Much uncertainty remains about the final spin of such systems. For example, a high spin value at core-collapse makes CHE systems exemplary gamma-ray burst (GRB) progenitors. The disk winds and jets produced during a GRB could reduce the final spin to become consistent with our observations \citep[e.g.,][]{Gottlieb_2024}. However, these same processes would also lead to significant mass loss, causing tension with the location in the mass distribution.
Hence it is challenging to reconcile the CHE channel with both the high masses and the low effective spins observed.
In terms of mass ratios, CHE systems are expected to form BBH systems with nearly equal masses, as efficient tidal locking requires similar-mass components. 
\citet{Hastings_2020} find that CHE systems have mass ratios $q \gtrsim 0.8$ (see their Fig.~8), which could be consistent with the high mass ratios observed in our sample.

% Variations on the CHE story
Several variations on the canonical CHE scenario have also been proposed. 
Most notably, \citet{Briel2023} find that the $35\Msun$ feature in \texttt{BPASS} simulations is caused by stable mass transfer, whose regime is bounded by mass transfer stability, quasi-homogeneous evolution, and stellar winds. We refer to this as the ``SMT + quasi-CHE'' scenario ({\bf 2.B} in Fig.~\ref{fig:astro table}). 
We show their predicted mass and mass ratio distributions in Fig.~\ref{fig:compare form channels} (light blue lines). 
Their prediction for the mass function below the peak is seemingly much flatter than our results. However, their predictions tentatively match the location of the peak and the \edit{drop-off} to higher masses. 
On the other hand, the mass ratios in their simulations are significantly lower than those observed in our sample.
% Combining CHE + PISN
Another recent explanation by \cite{Winch2025} attributes the $35\Msun$ feature as a combined result of ``avoiding'' pair pulsations and rapid rotating stars.
Similar to the ``CCSNe to P-PISNe boundary'' scenario explained above, they find that the maximum mass just \textit{before} pair pulsations sets in occurs at a CO core mass of $M_{\rm CO,crit} = 36.3 \Msun$, close to the observed GW feature. 
They argue this will be the final stellar mass for CHE stars, since these stars experience enhanced mechanical mass loss \citep[while lower-mass stars retain their envelope and could fill the PISN-mass gap][]{Winch_2024}.
However, their models omit the orbital widening caused by mass loss, which could inhibit gravitational-wave mergers within a Hubble time.

In conclusion, while CHE stars are expected to become relevant in the mass distribution of merging BHs around $35\Msun$, it is difficult to reconcile this channel (and variations) with the observed low effective spins.

%%%%%%%%%%%%%%%%%%%%%%%%%%%%%%%%
\subsection{Population III stars }
% Characteristics of the population III stars
\edit{Pop}~III stars (row {\bf 3} in Fig.~\ref{fig:astro table}) form more massive BHs due to higher initial masses and weaker radiation-driven stellar winds \citep[e.g.,][]{Kinugawa_2014, Kinugawa_2020}.
% \citet{Kinugawa_2014, Kinugawa_2020} highlight how two key differences between Pop III and Pop I/II stars.
% Firstly, Pop III stars form more massive BHs due to higher initial masses and weaker radiation-driven stellar winds.
% Second, pop III stars with ZAMS masses $< 50,\Msun$ avoid a the red supergiant phase and thus unstable mass transfer.
%
% How this would create a $35\Msun$ peak
Because of this, \edit{Pop}~III stars have been suggested as a plausible origin for the $35\Msun$ peak \citep{2016PTEP.2016j3E01K, Kinugawa_2020,2023arXiv231217491I}.
Specifically, \citet{Kinugawa_2020} use rapid population synthesis to show that Pop III stars produce a broad feature peaking around $30\Msun$ (yellow line in Fig.~\ref{fig:compare form channels}). Although this feature very roughly matches observations, their model generally over-predicts the rate at $z = 0$, and does not display the steep drop we infer towards higher masses. 
They explain that this feature arises from the characteristic evolution of Pop III stars with initial masses above and below $50\Msun$. 
Stars below this threshold avoid the red supergiant phase and undergo only stable mass transfer, losing $10-30\%$ of their mass in their models. 
In contrast, more massive stars experience common-envelope evolution and lose about $50-70\%$ of \edit{their} mass in their models. 
This conspires to produce a feature that peaks around $30\Msun$, regardless of several variations in the stellar evolution models \citep[Fig.~3 in][]{Kinugawa_2020}.
% What do the 'other properties' say
Pop III star formation peaks at very high redshift ($z \sim 7-10$), implying that binaries merging today must have experienced long delay times. 
This has important consequences for the spin: while short-delay mergers at high redshift may have undergone efficient tidal spin-up, the progenitors of present-day mergers likely avoided tidal interactions, resulting in low expected spins \citep{Kinugawa_2020}.
We show the effective spin distribution of \emph{detectable} Pop III stars as predicted by the fiducial model from \cite{Kinugawa_2020} (their Fig.~12) in the bottom left panel of Fig.~\ref{fig:compare form channels}
(it is a reasonable comparison since the selection function for positive $\chieff$ is not strong \citep{Ng_2018}).
\citet{2016PTEP.2016j3E01K} further find that Pop III mergers exhibit a broad mass-ratio distribution, which is consistent with our inferred results. 
Lastly, we expect \edit{Pop}~III stars to follow a formation history that differs from the overall cosmic star formation rate.
Currently, we find that the population in the $35\Msun$ peak is consistent with the overall star formation rate (see Section~\ref{sec:Result}), which would not argue in favor of a Pop~III star origin. However, current data limitations prevent us from drawing any firm conclusions and future high-redshift observations will be essential for a more robust test.

% \chcomment[id=WF]{Make sure to note all the X and checks in the table throughout this section.  We should also be careful to call out generic ``hierarchical formation'' and this section's GCs.} 

%%%%%%%%%%%%%%%%%%%%%%%%%%%%%%%%
\subsection{Globular clusters}

% How this would create a $35\Msun$ peak
Dynamical environments, such as young
stellar clusters, globular clusters, and nuclear star clusters, naturally favor the formation of more massive mergers, as dynamical interactions tend to pair the most massive systems. 
In this subsection, we focus on how the excess near $35\,\Msun$ could be the result of dynamical pairing in such environments. Signatures of hierarchical mergers (i.e., higher order generations) are discussed separately in Section~\ref{ss: hierarchical mergers} below.

Recent cluster population synthesis simulations by
\citet{Antonini_2023} show that globular clusters could indeed produce a peak
around $\sim 35\Msun$, provided that the most massive clusters form with
half-mass densities $\geq10^4\Msun/pc^{-3}$  (light and dark green lines in top panels of Fig.~\ref{fig:compare form channels}).
Their simulated globular cluster merger population broadly matches the features we observe in the primary mass distribution, both in terms of location and shape.
We use \citet{Antonini_2023} as our primary reference for drawing conclusions related to row {\bf 4}, ``globular clusters'', in Fig.~\ref{fig:astro table}.

% What do the 'other properties' say
In terms of mass ratio, their predictions appear to over predict the number of unequal-mass mergers compared to what we observe in the $35\Msun$ feature. 
This mismatch becomes more pronounced when we apply cuts in $\Mchirp$ rather than the fiducial $m_1$, $m_2$ selection (see appendix Figure \ref{fig: diff sel}).

\begin{figure*}
    \centering
    \includegraphics[width=0.48\textwidth]{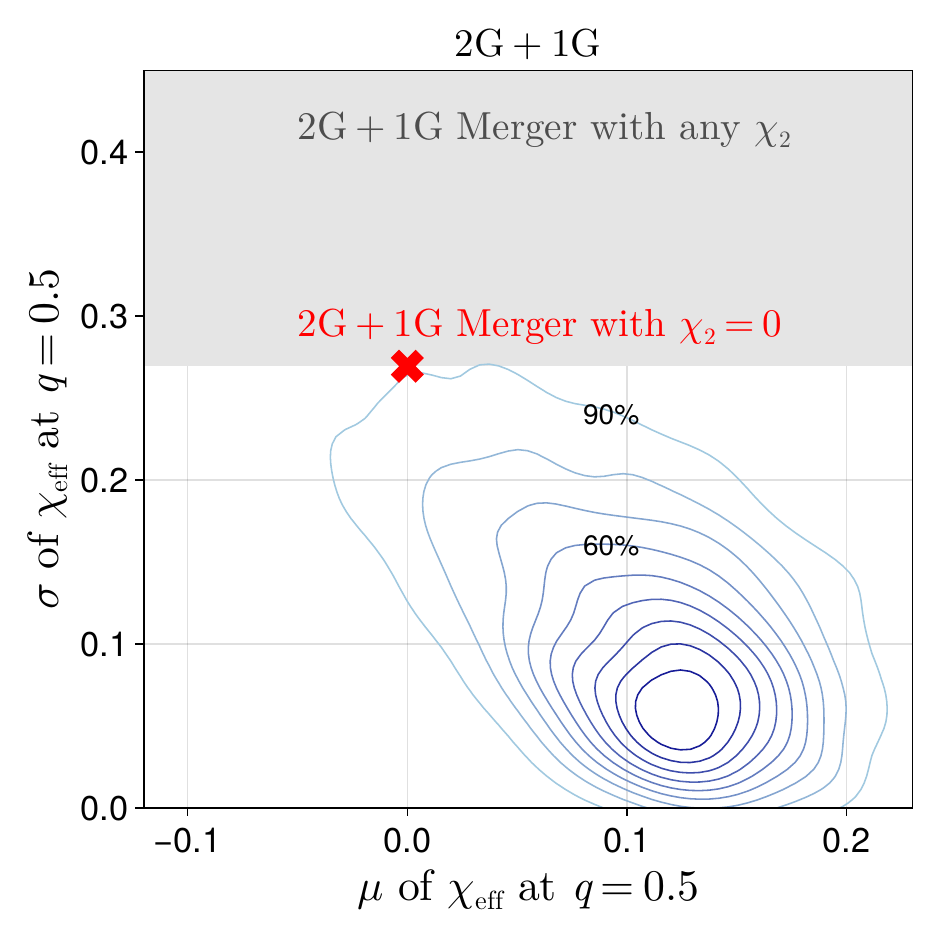}
    \includegraphics[width=0.48\textwidth]{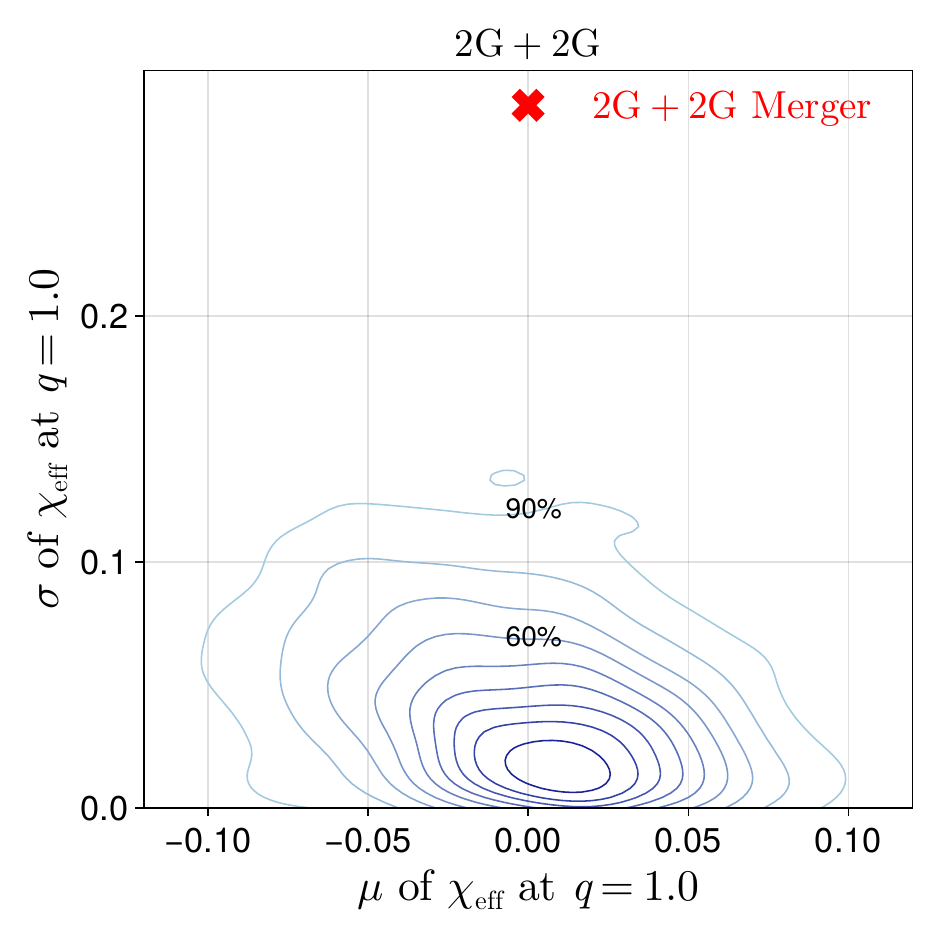}
    \caption{The mean and standard deviation of the effective spin distribution. 
    The predicted value for hierarchical mergers is shown in red, while our analysis for the $35\Msun$ peak population is shown in blue contours, which denote the 10\%, 20\%, \ldots, 90\% credible regions from our inference.
    This shows that the observed $q$ and $\chieff$ values are inconsistent with a hierarchical merger formation origin. 
    We find significant posterior support at $q=0.5$ and $q=1$ from reweighted PEs, indicating that our results are not based on extrapolation.
    See the \href{https://github.com/SoumendraRoy/35Msun_GWTC3/blob/main/m1m2_cut/scripts/Plots_Main.ipynb}{code} used to generate this plot.}
    \label{fig: hier mergers}
\end{figure*}

Lastly, dynamical formation scenarios predict isotropic spin–orbit orientations, which lead to an effective spin distribution symmetric around zero \citep[e.g.,][]{Rodriguez_2018}\footnote{Small positive values of $\chieff$ in BBH mergers can arise through accretion; see \cite{2025ApJ...983L...9K}.}. In contrast, we find that the $\chi_{\rm eff}$ distribution prefers positive values, rather than symmetric around zero (consistent with the findings of \cite{sadiq2025seekingspinningsubpopulationsblack} and the trend observed in the full GWTC-3 population \citep{gwtc-3pop}).
Interestingly, \citet{Ray2024} report that systems which contribute to the excess in $30-40\Msun$ range show tentative evidence for median $\chi_{\rm eff}$ of zero, which they interpret as a plausible signature of dynamical formation. 
However, they also note \edit{that} the large uncertainties in their analysis allow for multiple interpretations. 
Our results are statistically consistent with those of \citet{Ray2024}, despite the differing interpretations \edit{(see \ref{appen: compare chieff} for details)}. 
% (see Fig.~\ref{appen: different Mass Cut} in Appendi). 
Additional data will be necessary to reach more definitive conclusions.

We conclude that globular clusters remain a plausible formation channel for the $35\Msun$ feature, though notable tensions persist in the predicted mass ratio and effective spin distributions.

%%%%%%%%%%%%%%%%%%%%%%%%%%%%%%%%
\subsection{Hierarchical mergers \label{ss: hierarchical mergers}}

% \chcomment[id=LvS]{@SR and WF, let me know if you agree with the current structure.}
% \chcomment[id=WF]{Re-write a bit, noting mass conservation arguments,
% cite Vecchio (in prep) and Zoheyr, and argue why there are only ``even peaks''
% and not ``odd peaks'' in the mass function (not part of this analysis, but
% relevant). Then, finally, turn to $\chi_\mathrm{eff}$-$q$.}

% How this would create a $35\Msun$ peak
Several works have proposed that the $35\Msun$ feature could be caused by
hierarchical mergers which combine lower-mass BHs from peaks in the mass distribution at $9\Msun$ and (the tentatively detected) $15-20\Msun$ \citep{TiwariFairhurst2021, Mahapatra2025}. 
In this case, the primary BHs that contribute to the $35\Msun$ feature are second (2G), third (3G), or even fourth-generation mergers. 
Higher generation merger products are difficult to retain due to the high merger-kick recoil velocities \citep{Campanelli_2007a,Campanelli_2007b}, \edit{and} only the densest environments (like AGN disks) can retain such systems 
\citep[e.g.~see][and references therein]{ford2025usinggravitationalwavesmultimessenger}.
Because of this, we will focus the discussion below on the 2G+1G, and 2G+2G merger scenarios ({\bf5A} and {\bf5B} respectively in Fig.~\ref{fig:astro table}), where the first generation (1G) refers to the original black holes formed from stellar collapse, and the second generation (2G) refers to black holes formed from mergers of 1G black holes.

Hierarchical mergers make a particularly distinct prediction for the effective spin distribution due to the combination of two effects \citep[see also e.g.,][]{Antonini2025}.
First, the spin magnitudes of second (or higher) generation merger products are strongly peaked around $\sim 0.7$ \citep[e.g.,][]{Pretorius_2005, Buonanno_2008, Hofmann2016, Fishbach2017, Gerosa_2017}.
Second, since the black holes are dynamically paired, the orientation between the spins and orbital angular momentum is isotropically distributed \edit{in absence of sufficient gas-accretion \citep{2025ApJ...983L...9K}}.
As a result, the mean of the effective spin distribution is expected to be zero.
We can furthermore calculate the standard deviation of the $\chieff$ distribution as follows:
\begin{equation}
    \sigma(\chieff) = \frac{1}{(1+q)}[\mathrm{var}(\chi_1 \cos\theta_1) + q^2~\mathrm{var}(\chi_2 \cos\theta_2)]^{1/2}.
    \label{eq:var chieff}
\end{equation}
Here, $q$ is the mass ratio defined as $m_2/m_1$, $\chi_1$ and $\chi_2$ are the spin magnitudes of the individual black holes, and $\theta_1$ and $\theta_2$ are the angles between the spins and the orbital angular momentum.
Since these black holes are dynamically assembled, $\cos\theta_1$ and $\cos\theta_2$ follow uniform distributions over $[-1,1]$, and we assume zero covariance between $\chi_1\cos\theta_1$ and $\chi_2\cos\theta_2$.

%%%%%%%%%%%%%%%%%%%
\subsubsection*{2G + 1G mergers:}
In the case of a 2G+1G merger, the mass ratio is typically $q \approx 0.5$.
This is inconsistent with our results for the $35\Msun$ population, which tends to favor equal-mass systems.
Nevertheless, we can use Eq.~\ref{eq:var chieff} to calculate the expected variance of $\chieff$.
The more massive BH is second-generation, hence has spin magnitude of $\chi_1 \approx 0.7$. The variance $\mathrm{var}(\chi_2 \cos\theta_2)$ is always greater than zero, regardless of the value of $\chi_2$ and $\cos \theta_2$.
It sets a lower limit on the standard deviation of $\chieff$: $\mathrm{std}(\chieff) > 0.27$ for any population of 2G + 1G mergers.
We show this lower limit in the left panel of Fig.~\ref{fig: hier mergers}, where we compare the expected mean and standard deviation for $\chieff$ at $q=0.5$ to the inferred values for the $35\Msun$ population.
This clearly shows that the mean and standard deviation for 2G + 1G mergers fall outside the $90\%$ credible interval of the observed data, indicating a mismatch in both the mean and the spread of the $\chieff$ distribution.

%%%%%%%%%%%%%%%%%%%
\subsubsection*{2G + 2G mergers\edit{:}}
If both black holes are second-generation (2G), the expected mass ratio is $q \approx 1$, which is more consistent with our findings for the $35\,\Msun$ population. 
% Sort of mass conservation argument
However, if 2G + 2G mergers significantly contribute to the $35\,\Msun$ peak, we
should also observe a population of 2G + 1G systems \citep{Doctor2020}. Given
the larger recoil kicks of 2G BHs, the 2G + 1G population would be even more
prominent than the former.

% Expected variance of chieff
Regardless, we can again use Eq.~\ref{eq:var chieff} to calculate the expected variance of $\chieff$ for 2G + 2G mergers.
We now expect both spin magnitudes $\chi_1 \approx \chi_2 \approx 0.7$. 
Under these assumptions, the expected standard deviation of $\chieff$ is approximately $0.28$, as shown in the right panel of Fig.~\ref{fig: hier mergers}.
This prediction is around $6\sigma$ away from the inferred values at $q=1$.

In reality, $\chi_1$ and $\chi_2$ follow distributions centered around $0.7$ rather than exactly $0.7$ \citep{Gerosa_2017, Fishbach2017, borchers2025gravitationalwavekicksimpactspins}, which would only increase the spread in $\chieff$. This further strengthens the conclusion that hierarchical mergers \edit{in absence of sufficient gas accretion} are unlikely to account for the observed properties of the $35\Msun$ population. The same logic applies to even higher-generation mergers. \\

In conclusion, we find that the effective spin distribution of the $35\Msun$ peak population is not consistent with the predictions of hierarchical mergers, regardless of whether we consider 2G+1G or 2G+2G mergers.
Moreover, the observed mass ratio distribution is also inconsistent with the predictions of 2G+1G mergers.
While the mass ratio could be consistent with 2G+2G mergers, the absence of a significant population of 2G+1G mergers challenges the interpretation that 2G+2G mergers are a dominant contributor to the $35\Msun$ peak.
Therefore, we conclude that the $35\Msun$ peak cannot originate from the hierarchical merger channel.

%%%%%%%%%%%%%%%%%%%%%%%%%%%%%%%%%%%%%%%%%%%%%%%%%%%%%%%%%%%%%%%%%%%%%%%%%%%%%%%%%%%%%%%%%%%%%%%%
\section{Conclusion and Summary}\label{sec:conclusion}

\begin{figure*}
    \centering
    \includegraphics[width=\linewidth]{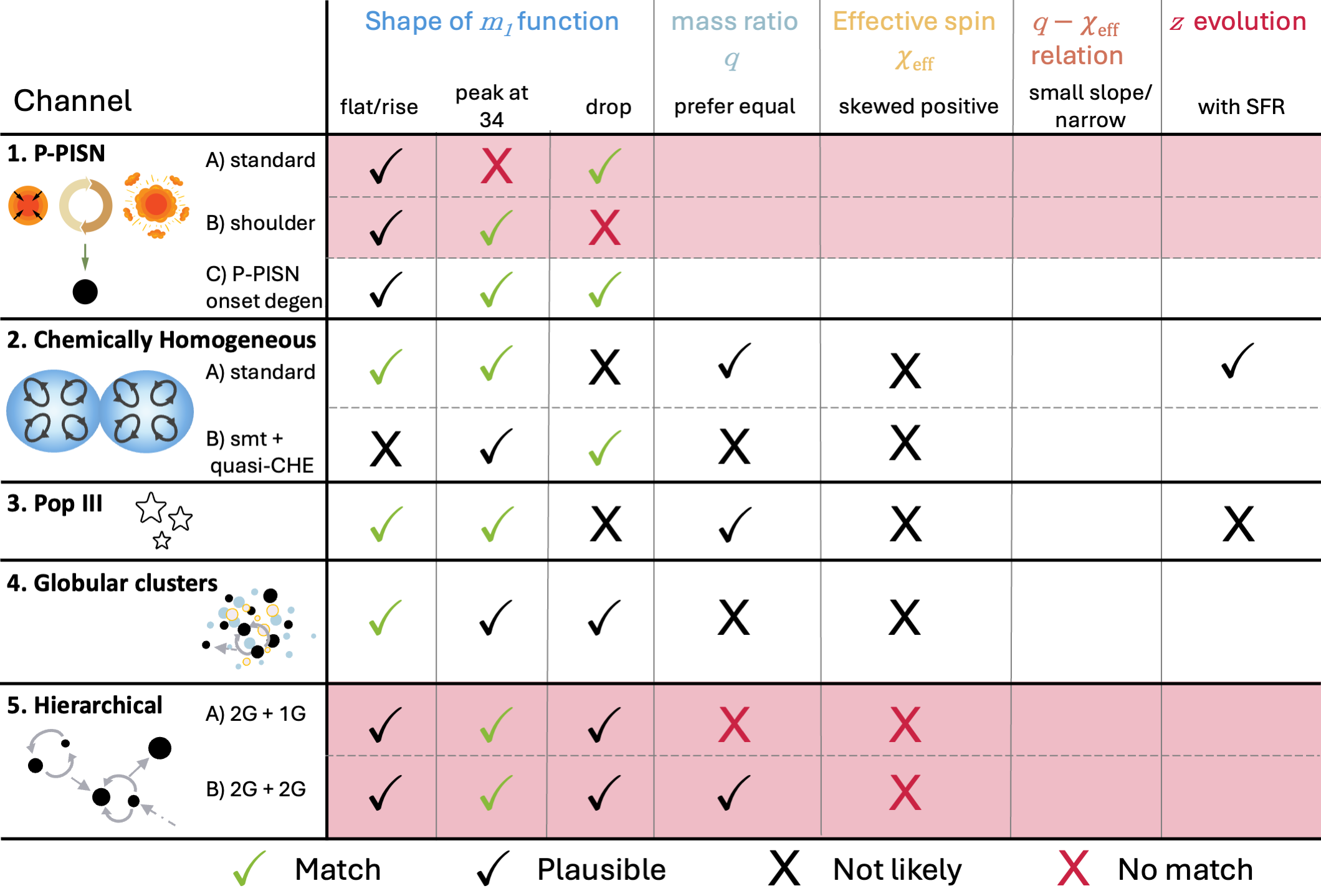} 
    \caption{Summary of our assessment of whether various formation channels are consistent with the inferred properties of the $35\Msun$ peak population.
    The cartoon for the P-PISN channel is adapted from \cite{Farag2022}.
    We evaluate the properties as follows:
    Match indicates robust agreement with observations. Plausible means the feature matches, though the robustness is uncertain. Not likely implies disagreement, but with potential flexibility in the models to agree. No match denotes robust inconsistency with the observed feature.  We find a ``mid-thirties crisis'' because no model is able to account for all the features we observe in the $35 \, \Msun$ peak.
    }
    \label{fig:astro table} % Add a label for referencing
\end{figure*}

% \chcomment[id=WF]{We need to \LaTeX{} this table.} --< \chcomment[id=LvS]{you know we never will...} 
 
% In this work, we have shown how isolating a specific feature in the mass distribution to investigate the `other' properties of this feature can provide insights into the formation channels of binary black holes.
% By doing this we help identify the common properties of this potential sub population,

In this work, we have isolated BBH mergers near the $35\Msun$
feature in the primary mass distribution, in order to analyze them separately from the rest of the population. 
This allows us to examine the distributions of primary mass, mass ratio, effective spin and redshift for systems contributing to this peak population. 
% With the goal of distinguishing between different proposed formation channels, 
We set out to address five specific questions outlined in the introduction.
Our main conclusions are as follows:

\paragraph{1.~The shape of the mass function}
    We find the BBH merger rate follows \edit{a} flat or shallow rise with $m_1$, increasing by a factor of 3 from $20\Msun$ up to a peak at $\MBHmax=33.9^{+3.7}_{-4.2}\Msun$ (see top left panel of Fig.~\ref{fig: main results}). 
    This is followed by a steep drop (decreasing by an order of magnitude between the peak and $50\Msun$).

\paragraph{2.~Mass ratio distribution}
    The systems within the peak population tend to favor equal-mass binaries over unequal-mass ones (top right panel of Fig.~\ref{fig: main results}). 
    This preference for equal mass systems is stronger in the $35\Msun$ peak population with respect to the full population of BBHs when similar mass function and pairing are used \citep{Golomb2024}. Compared to the canonical LVK models, which have structure in $m_1$ but not in $q$ the full population prefers equal mass mergers more strongly than our model \citep{gwtc-3pop}. 
    Removing GW190412 results in a much stronger preference for equal mass mergers in our model.
    
\paragraph{3.~Effective spin distribution}
    The distribution of effective spins of the peak population is skewed toward positive values, with negative median spin excluded at around $90\%$ confidence level (bottom left panel of Figure \ref{fig: main results}).
    This is consistent with the effective spin distribution inferred for the overall BBH merger population by \cite{gwtc-3pop}.
    %This result is not very sensitive to the prior selection cut in the BBH masses that was used (see Fig.~\ref{fig: diff sel}). 

\paragraph{4.~Effective spin -- mass ratio distribution}
    The presence of a correlation is unconfirmed in our analysis. 
    We find both $\alpha_{\chieff}$ (which measures the change of the mean $\chieff$ with $q$), and $\beta_{\chieff}$ (which measures the change in the width of the $\chieff$ distribution with $q$) are consistent with zero (Fig.~\ref{fig: q-chieff pop}). 
    We do note that most of the posterior support lies at $\alpha_{\chieff} < 0$, hinting at the same anti-correlation observed for all BBH events in GWTC-3 \citep{gwtc-3pop,Callister2021}.

\paragraph{5.~Rate evolution with redshift}
    The redshift evolution of the rate is consistent with the shape of the star formation rate (bottom right panel of Fig.~\ref{fig: main results}), with decreasing or constant rates clearly ruled out. Notably, the $35\,\Msun$ population provides tighter constraints on the rate evolution than the full BBH population, suggesting it carries most of the constraining power.

%%%%%%%%%%%%%%%%%%%%%%%%%%%%%%%%
\subsection*{What is the astrophysical origin of the $35\Msun$ peak?}

We systematically compare our findings for the $35\Msun$ peak population to a range of proposed formation channels in Section~\ref{sec:astrointerp}.
To the best of our ability, we summarize these comparisons in the table presented in Fig.~\ref{fig:astro table}.
While many studies provide predictions for the mass distribution, few offer quantitative forecasts for additional properties such as the mass ratio, effective spin, and redshift.
Fig.~\ref{fig:astro table} reveals what we refer to as a “mid-thirties crisis”: none of the currently proposed channels satisfactorily checks all the boxes of the observed features. 

% This reveals mid-thirties crisis: none of the currently proposed channels check all the boxes. 

Due to the flexibility in many of the predictions associated with different formation channels, it is challenging to definitively rule out any given scenario.
However, based on our analysis, we identify the following channels (highlighted in red in Fig.~\ref{fig:astro table}) as highly unlikely to explain the observed features of the $35\Msun$ peak population, due to robust inconsistencies with at least one key property (indicated by red crosses in the corresponding table entries).

The standard P-PISN scenario ({\bf 1A})  appears unlikely for several reasons (see Section~\ref{ss: PPISN}), but the most compelling is \edit{that} the location of the PISN mass gap cannot be brought down to the observed $35\Msun$ range. 
Current models place the lower edge of the gap no lower than $\sim45\Msun$,  whereas we confidently find the peak of the observed distribution at $33.9^{+3.7}_{-4.2},\Msun$.
The P-PISN shoulder scenario ({\bf 1B}) is highly unlikely because here, the primary P-PISN peak is expected to occur at higher black hole masses (around $45\,\Msun$), originating from more common lower-mass stellar progenitors. As a result, the mass function should rise beyond the $35\Msun$ peak, leading to a second, higher-rate peak at $\sim45\Msun$. However, our analysis confidently excludes any such rise beyond $35\Msun$.
Lastly, the hierarchical merger scenario \edit{in absence of sufficient accretion} {\bf 5} (including both 2G+1G and 2G+2G combinations) is robustly excluded due to a mismatch between the observed properties and the predicted mass ratio and effective spin distributions (see Section~\ref{ss: hierarchical mergers}).

\edit{In principle, the uncertainties of binary population synthesis, such as initial mass function (IMF) of massive progenitors, the mass-ratio and orbital period distribution, and the physics of binary interactions, could affect the observed $35\,\Msun$ peak. For example, a top-heavy IMF increases the formation efficiency of massive black holes. The mass-ratio and orbital period distributions contribute to the stability and timescale of binary evolution, hence deciding which black holes will merge within Hubble time. Similarly, assumptions about mass-transfer stability and common-envelope evolution affect how efficiently binaries harden, setting the merger efficiency across the mass spectrum. The natal kicks from supernovae preferentially disrupt lower-mass systems and increase the rate of massive binaries. However, while these effects indeed affect the overall rates of BBH mergers, they do not produce a pile-up at $\sim 35\, M_{\odot}$, at least by any known physical phenomena.}\\

A detailed multidimensional analysis of the $35 \, \Msun$ peak became possible by restricting our study to BBH merger events near the peak. We see such targeted population analyses as a promising direction for GW population studies. Since event selection with incomplete population model can introduce bias, we apply mass cuts where PE uncertainties are well below the separation between features identified by non-parametric reconstructions \citep[cf.][]{autoreg1}. We test robustness by varying these cuts across multiple choices (see Fig.~\ref{fig: diff sel}) and will detail the methodology in a forthcoming paper. \\
 
With the ongoing fourth GW observing run set to quadruple the number of detections \citep{gracedb}, the kind of multidimensional analysis presented here will become increasingly constraining.  We eagerly anticipate the fifth observing run featuring two high-sensitivity LIGO detectors in the United States, with Virgo, KAGRA, and LIGO-India joining the network. This expanded configuration will quadruple the detected BBH population again, allowing for much tighter constraints compared to those presented in this work. 
Combining mass, mass ratio, spin, and redshift, especially when focusing on a particular feature of merging \edit{populations} as we do here, offers a powerful path to uncover subpopulations of GW sources and their formation channels. 
To make full use of this approach, we urgently need more complete and holistic predictions for formation channels (many entries in Fig.~\ref{fig:astro table} remain empty). 
We encourage the astrophysical modeling community to address these gaps.

%%%%%%%%%%%%%%%%%%%%%%%%%%%%%%%%%%%%%%%%%%%%%%%%%%%%%%%%%%%%%%%%%%%%%%%%%%%%%%%%%%%%%%%%
\section{Data Availability}
The code is publicly available on  GitHub: \url{https://github.com/SoumendraRoy/35Msun_GWTC3} under the GNU General Public License. A frozen version of the code, along with the data produced by this work, is also available on Zenodo: \href{https://doi.org/10.5281/zenodo.15778343}{10.5281/zenodo.15778343}.

%%%%%%%%%%%%%%%%%%%%%%%%%%%%%%%%%%%%%%%%%%%%%%%%%%%%%%%%%%%%%%%%%%%%%%%%%%%%%%%%%%%%%%%%
\section{Software}
\edit{\texttt{Turing.jl} \citep{ge2018t}; \texttt{Makie.jl} \citep{DanischKrumbiegel2021}; \texttt{PopModel.jl} \citep{popmodel}.}

%%%%%%%%%%%%%%%%%%%%%%%%%%%%%%%%%%%%%%%%%%%%%%%%%%%%%%%%%%%%%%%%%%%%%%%%%%%%%%%%%%%%%%%%
\section{Acknowledgment} 

We thank Maya Fishbach for the LIGO P\&P review. We thank Lucas de Sa for sharing the result on CHE channel, Vishal Baibhav and Aditya Vijaykumar for the discussions on hierarchical mergers, Maya Fishbach and Thomas Dent for suggesting the inclusion of GW190412 in the selected catalog, and Tomoya Kinugawa for suggesting us the references of Pop III channel. We are grateful to Thomas Dent and Anarya Ray for sharing the data from \cite{sadiq2025seekingspinningsubpopulationsblack} and \cite{Ray2024} respectively. We are also thankful to the CCA GW group and LVK R\&P group for many stimulating discussions.

SKR thanks the Center for Computational Astrophysics at the Flatiron Institute for hospitality while this research was carried out. The computations in this work were, in part, run at facilities supported by the Scientific Computing Core at the Flatiron Institute, a division of the Simons Foundation. SKR also thanks the Institute for Cosmic Ray Research of the University of Tokyo, and especially Soichiro Morisaki, for their hospitality during the period when part of this work was carried out.

This material is based upon work supported by NSF’s LIGO Laboratory which is a major facility fully funded by the National Science Foundation. This research has made use of data or software obtained from the Gravitational Wave Open Science Center (gw-openscience.org), a service of LIGO Laboratory, the LIGO Scientific Collaboration, the Virgo Collaboration, and KAGRA. LIGO Laboratory and Advanced LIGO are funded by the United States National Science Foundation (NSF) as well as the Science and Technology Facilities Council (STFC) of the United Kingdom, the Max-Planck-Society (MPS), and the State of Niedersachsen/Germany for support of the construction of Advanced LIGO and construction and operation of the GEO600 detector. Additional support for Advanced LIGO was provided by the Australian Research Council. Virgo is funded, through the European Gravitational Observatory (EGO), by the French Centre National de Recherche Scientifique (CNRS), the Italian Istituto Nazionale di Fisica Nucleare (INFN) and the Dutch Nikhef, with contributions by institutions from Belgium, Germany, Greece, Hungary, Ireland, Japan, Monaco, Poland, Portugal, Spain. The construction and operation of KAGRA are funded by Ministry of Education, Culture, Sports, Science and Technology (MEXT), and Japan Society for the Promotion of Science (JSPS), National Research Foundation (NRF) and Ministry of Science and ICT (MSIT) in Korea, Academia Sinica (AS) and the Ministry of Science and Technology (MoST) in Taiwan. 

This paper carries LIGO document number LIGO-P2500403.

%%%%%%%%%%%%%%%%%%%%%%%%%%%%%%%%%%%%%%%%%%%%%%%%%%%%%%%%%%%%%%%%%%%%%%%%%%%%%%%%%%%%%%%%%%%%%%%%
\appendix

% \chcomment[id=WF]{Corner plot goes to appendix as well---maybe in landscape orientation.  Also show $R_0$, $\lambda$, $\alpha_{HM}$, $r_{HM}$, $\beta$, $\alpha_eff$, and $\beta_eff$, and $\mu_eff$ and $\sigma_eff$.  But if you need to cut it down, pick irrelevant variables.}

%%%%%%%%%%%%%%%%%%%%%%%%%%%%%%%%%%%%%%%%%%%%%%%%%%%%%%%%%%%%%%%%%%%%%%%%%%%%%%%%%%%%%%%%%%%%%%%%
\section{Results with Different Event Selection}\label{appen: different Mass Cut}

\begin{figure*}
    \centering

    \includegraphics[width=0.9\textwidth]{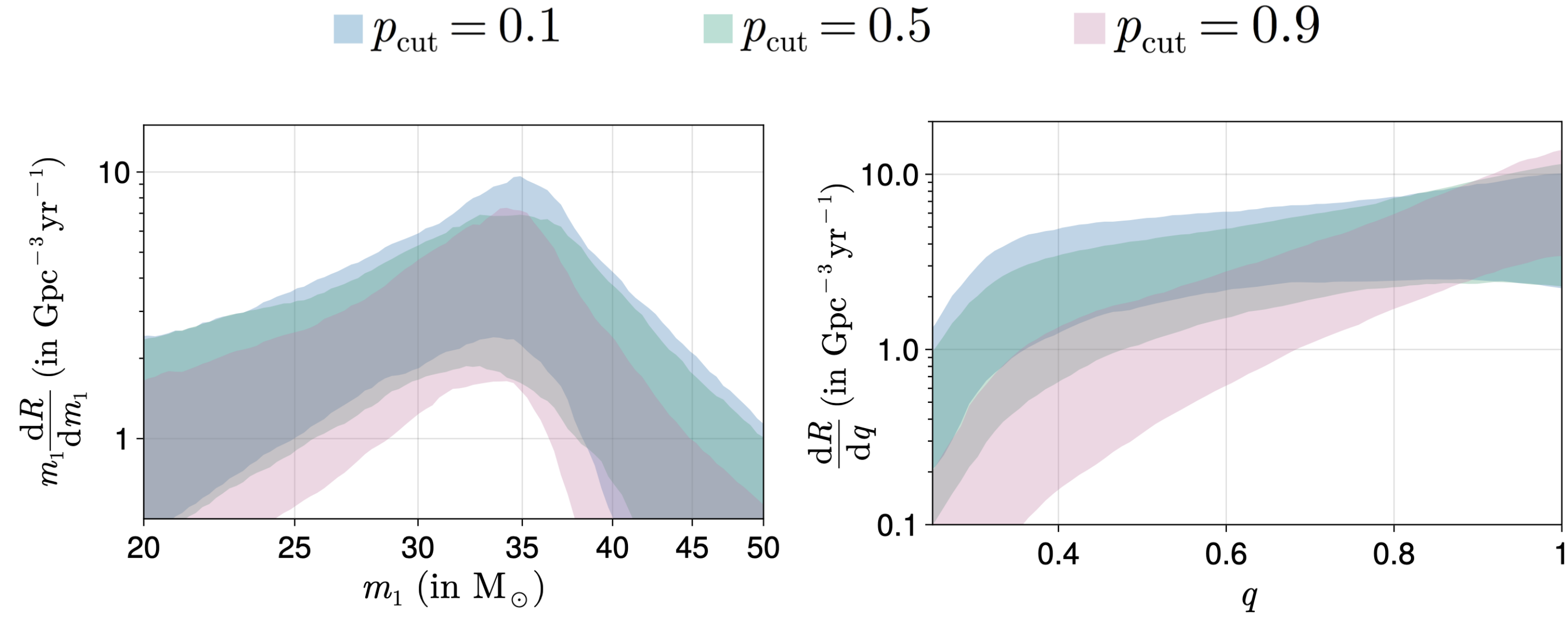}

    \includegraphics[width=0.9\textwidth]{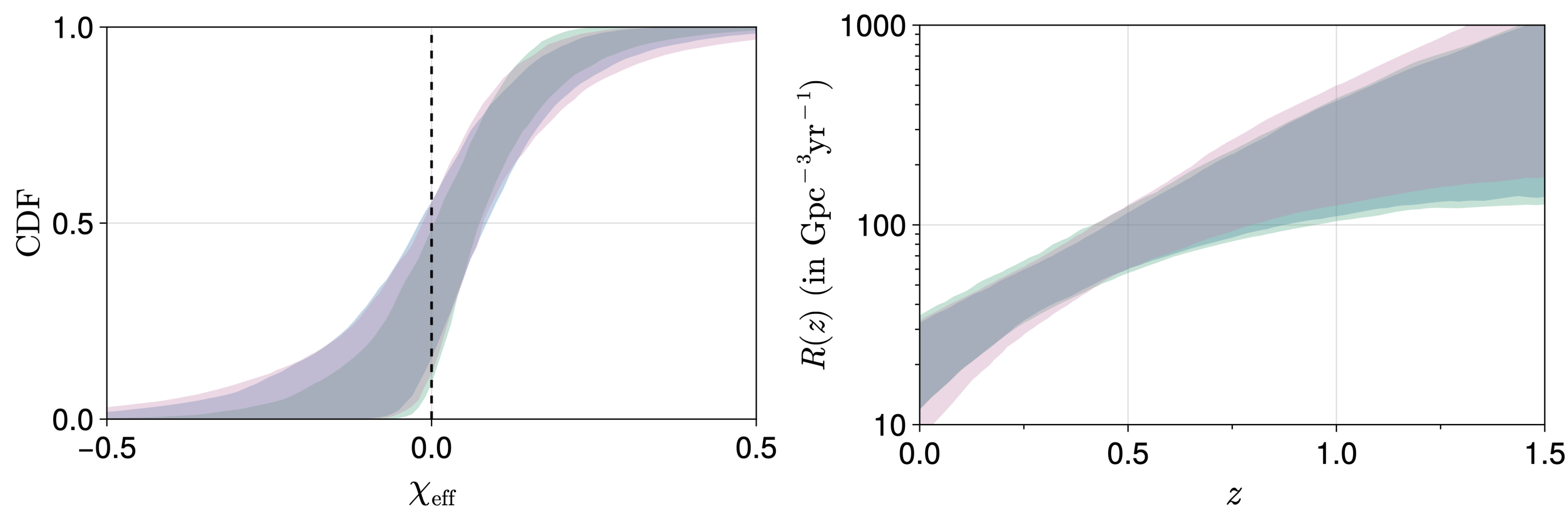}

    \vspace{0.5cm}
    
    \includegraphics[width=0.9\textwidth]{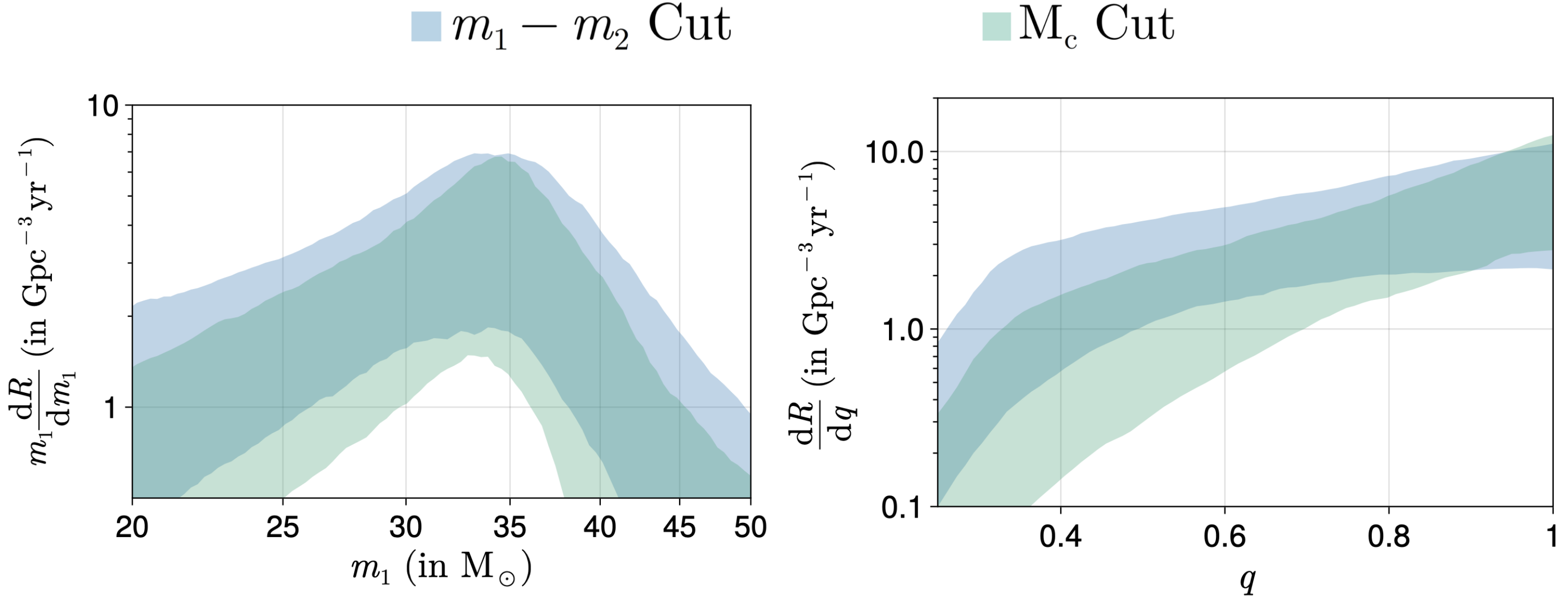}

    \includegraphics[width=0.9\textwidth]{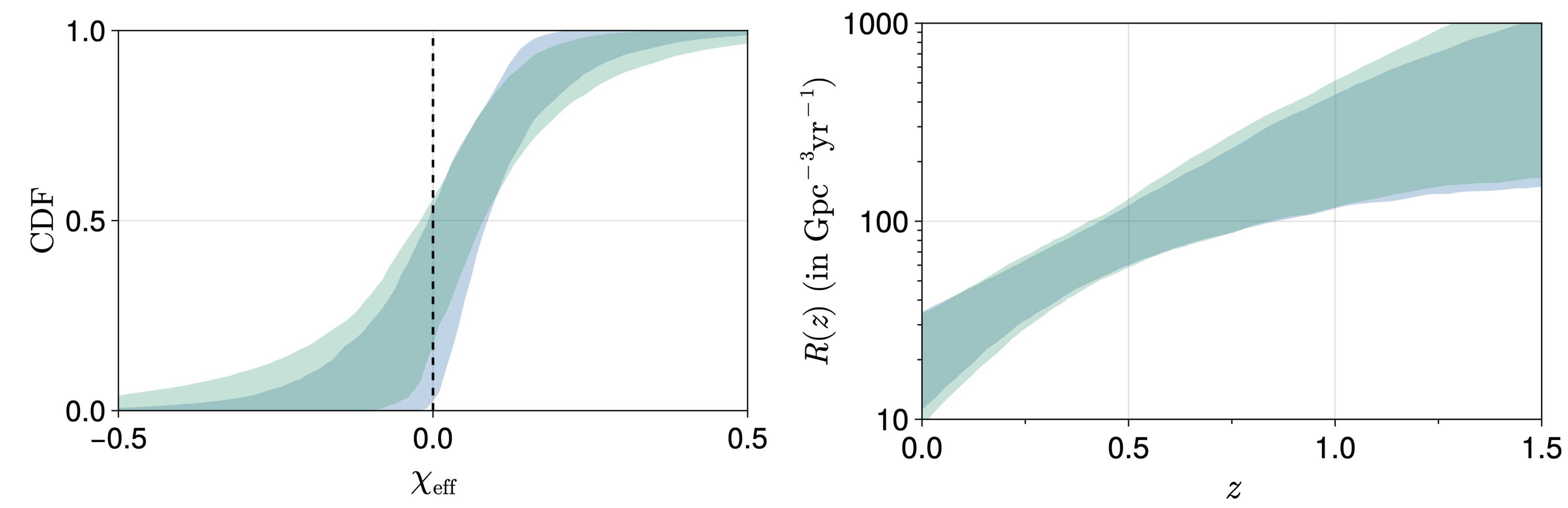}
    
    \caption{Results for different mass cuts. The top two panels show results for varying selection thresholds, with $p_\mathrm{cut}$ set to $0.1$, $0.5$ and $0.9$. The bottom panel focuses on events with source-frame chirp masses between those of $20-20\Msun$ and $50-50\Msun$ binaries, using $p_\mathrm{cut}=0.5$. We find that, apart from the mass ratio, the results remain essentially unchanged across different selection cuts. See the \href{https://github.com/SoumendraRoy/35Msun_GWTC3/blob/main/m1m2_cut/scripts/Plots_Appendix.ipynb}{code} used to generate this plot.}
    \label{fig: diff sel}
\end{figure*}

To assess the robustness of our catalog, we test the impact of varying two selection criteria: the threshold $p_\mathrm{cut}$ and mass cut. We consider $p_\mathrm{cut} = 0.1, 0.5, 0.9$. We find that the distributions of primary mass, redshift, and effective spin remain essentially unchanged across these values. However, the mass-ratio distribution becomes steeper as $p_\mathrm{cut}$ increases. The reason is that higher $p_\mathrm{cut}$ thresholds exclude systems with more extreme mass ratios, naturally steepening the inferred distribution.

For the mass cut, we select GWTC-3 events whose source-frame chirp mass falls between the chirp masses corresponding to equal-mass binaries of $20 \, \Msun$ and $50 \, \Msun$, with at least $50\%$ posterior probability. We again find that the distributions of primary mass, effective spin and redshift remain consistent under both selection cuts. This selection also reinforces our preference for equal-mass binaries as it excludes some of the more extreme mass-ratio systems, such as GW190412. We find weak ($2\sigma$) evidence for $\chi_\mathrm{eff}$ broadening with mass ratio $q$ in this subset of events.

%%%%%%%%%%%%%%%%%%%%%%%%%%%%%%%%%%%%%%%%%%%%%%%%%%%%%%%%%%%%%%%%%%%%%%%%%%%%%%%%%%%%%%%%%%%%%%%%

\section{Comparison of Effective Spin Populations with Ray et al.~(2024) and Sadiq et al.~(2025):}\label{appen: compare chieff}

\begin{figure*}
    \centering

    \includegraphics[width=0.495\textwidth]{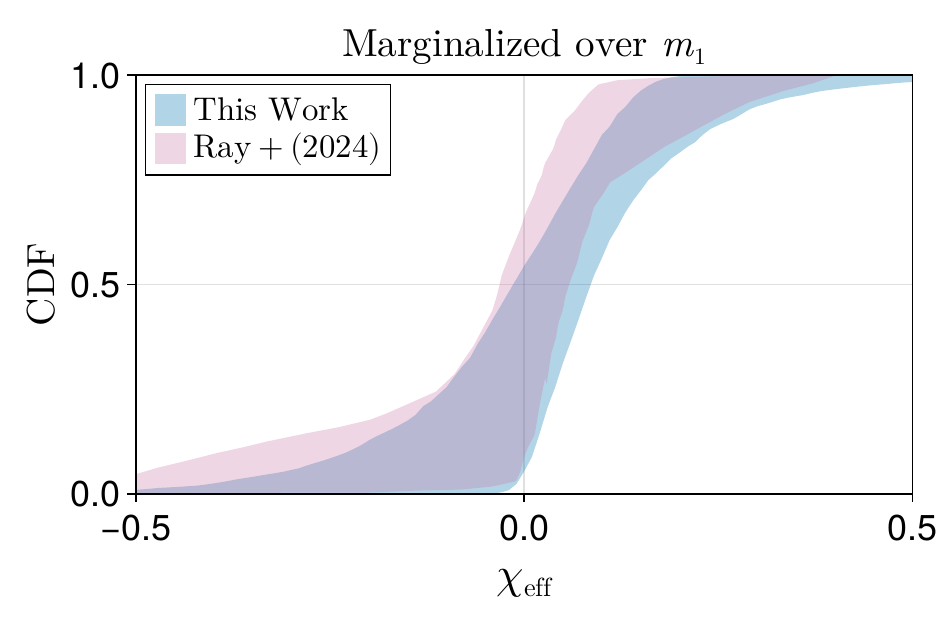}
    \includegraphics[width=0.495\textwidth]{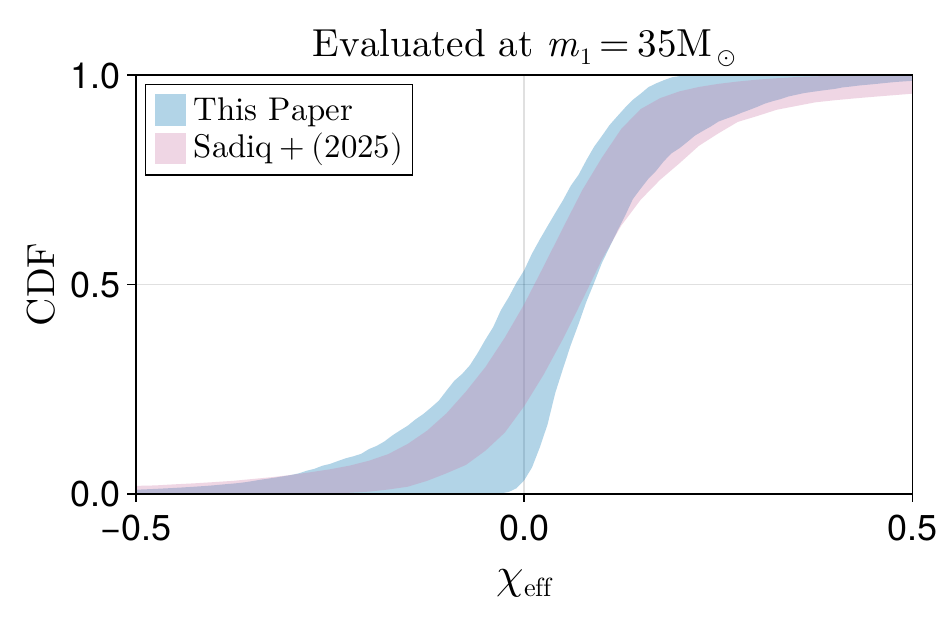}
    
    \caption{Comparison of $\chieff$ population with \cite{Ray2024} (left panel) and \cite{sadiq2025seekingspinningsubpopulationsblack} (right panel). \cite{Ray2024} find a median $\chieff$ value of zero for the $35\Msun$ peak population, which differs from our result. However, due to the large uncertainties, our distribution remains statistically consistent with theirs. Our $\chieff$ distribution is entirely consistent with the findings of \cite{sadiq2025seekingspinningsubpopulationsblack}. See the \href{https://github.com/SoumendraRoy/35Msun_GWTC3/blob/main/m1m2_cut/scripts/Plots_Appendix.ipynb}{code} used to generate this plot.
    }
    \label{fig: comp chieff}
\end{figure*}

In Fig.~\ref{fig: comp chieff}, we compare our marginalized $\chieff$ distribution with those reported by \cite{Ray2024} and \cite{sadiq2025seekingspinningsubpopulationsblack}. \cite{Ray2024} model the joint distribution of $m_1$, $m_2$, and $\chieff$ using the binned Gaussian process across all BBH mergers in GWTC-3. For the $30-40\Msun$ mass bin, they find the $\chieff$ distribution to be symmetric around zero. In the left panel of Fig.~\ref{fig: comp chieff}, our results are broadly consistent with theirs, primarily due to significant uncertainties. However, we find that a median $\chieff$ of zero is disfavored at approximately $90\%$ confidence.

\cite{sadiq2025seekingspinningsubpopulationsblack} evaluate the $\chieff$ distribution at $m_1=35\Msun$ and report a median significantly away from zero. Our result closely matches this finding, as shown in the right panel of Fig.~\ref{fig: comp chieff}.

%%%%%%%%%%%%%%%%%%%%%%%%%%%%%%%%%%%%%%%%%%%%%%%%%%%%%%%%%%%%%%%%%%%%%%%%%%%%%%%%%%%%%%%%%%%%%%%%
\section{Mock PE Set-up}\label{appen: mock PE}
We use the mock injections for GWTC-3 provided by \cite{gwtc-3inj} and identify those with a FAR below 1 per year. For each injection that crosses this threshold, we estimate the PE posterior by first calculating the signal-to-noise ratio (SNR). We do it by interpolating the injections on a grid of masses and rescaling them according to the luminosity distance. To simulate observational uncertainty, we add random Gaussian noise (with a standard deviation of 1) to the actual SNR to obtain the observed SNR.

Next, we estimate the uncertainties in the detector-frame masses by assuming they scale inversely with the observed SNR. We calibrate the proportionality constant by matching the standard deviation of the source-frame chirp mass from the catalog PEs. This calibration is consistent with the results of \cite{Vitale_2017}. In this way, we simulate the PE likelihood for each injection.

To generate posterior samples of detector-frame masses and luminosity distance, we adopt uniform priors for the detector-frame masses and a $D_L^2$ prior to the luminosity distance $D_L$. Full details of this procedure, including the mathematical formulation, are provided in Appendix C of \cite{Roy_2025}.

\section{Injection-Recovery Campaign}\label{appen: inj-rec}

\begin{figure*}
    \centering

    \includegraphics[width=\textwidth]{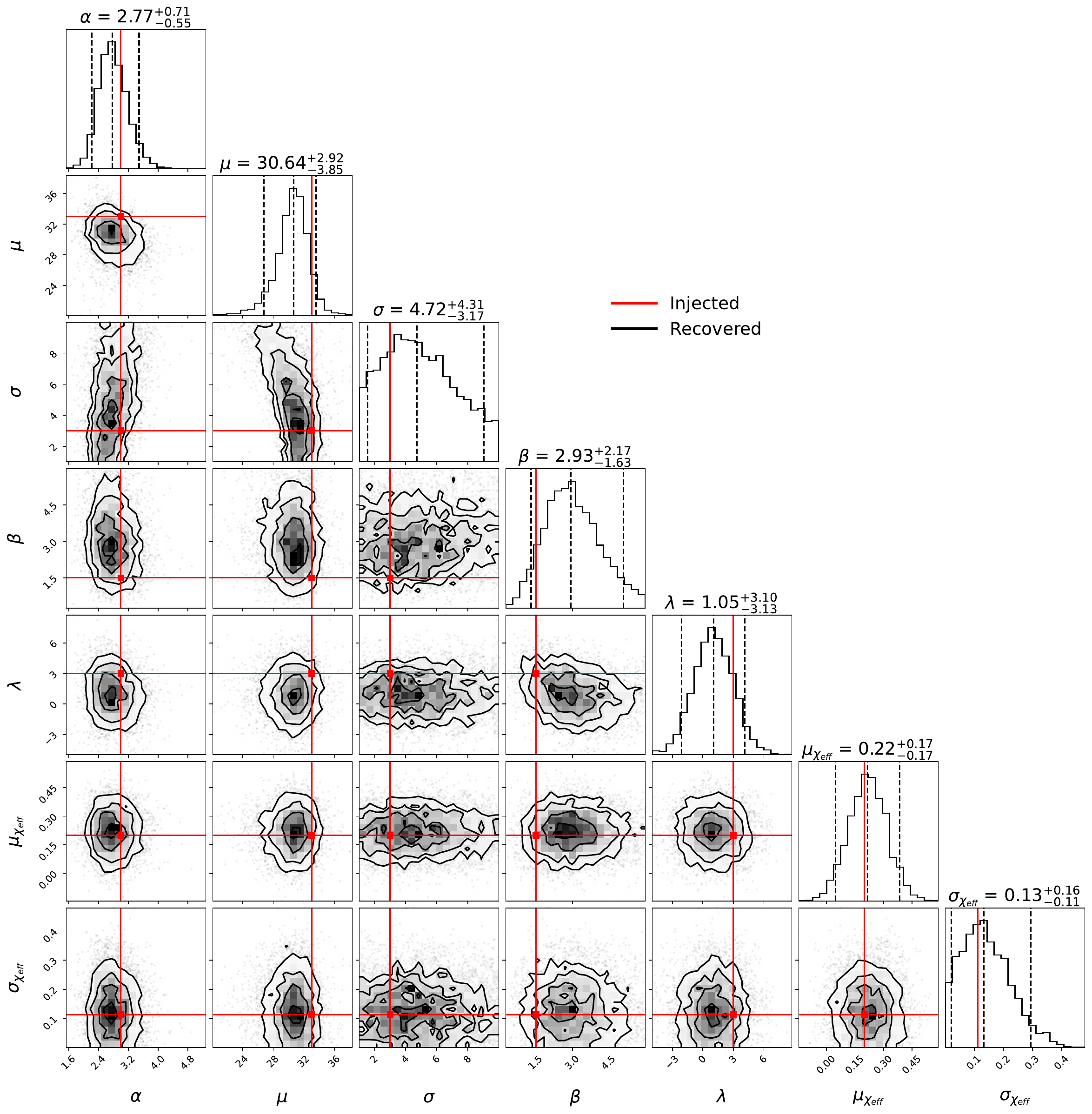}
    
    \caption{Injection–recovery campaign for the $m_1 - q - z - \chi_{\mathrm{eff}}$ population around the $35\,\Msun$ peak. We simulate $10,000$ events from the full population, and successfully recover the injected value after applying a cut to the observed space that isolates the $35\,\Msun$ peak. See the \href{https://github.com/SoumendraRoy/35Msun_GWTC3/blob/main/cutpop/m1m2_cut/scripts/plot.ipynb}{code} used to generate this plot.}
    \label{fig: inj rec}
\end{figure*}

To establish our method for selecting a subset of the catalog by applying cuts to the PE samples, we perform an end-to-end injection–recovery campaign using a dataset with a size comparable to the GWTC-3 catalog. We first assume the injected population distribution of $m_1$ to be:
 
\begin{equation*}
m_1 \sim 0.2\,N(m_1|\mu^{\prime}=10, \sigma^{\prime}=1.5)
\end{equation*}
\begin{equation}
+ 0.1\,N(m_1|\mu=33, \sigma=3)
\end{equation}
\begin{equation*}
+ 0.7\,\text{PowerLaw}(m_1|\alpha=3, m_{\text{min}}=3, m_{\text{max}}=150).
\end{equation*}

For the mass ratio $q$ and redshift $z$, we assume power-law distributions:

\begin{equation*}
q \sim \text{PowerLaw}(q|\beta, q_{\text{min}}, q_{\text{max}})
\end{equation*}
\begin{equation}
z \sim \text{PowerLaw}(1+z|\lambda, z_{\text{min}}, z_{\text{max}}),
\end{equation}

with $\beta=1.5,\, q_{\text{min}}=0,\, q_{\text{max}}=1,\,\lambda=3,\,z_{\text{min}}=0,\,z_{\text{max}}=2$.

For the effective spin $\chieff$, we assume a Gaussian distribution with mean $\mu_{\chieff}=0.2$ and standard deviation $\sigma_{\chieff}=10^{-0.95}$:

\begin{equation}
\chieff \sim N(\chieff|\mu_{\chieff}, \sigma_{\chieff}).
\end{equation}
We generate $69$ events from this population distribution with a SNR cut of $8$ and inject them into the GWTC-3 LVK noise. We then use \texttt{Bilby} \citep{Ashton2019bilby, Shaw2020BILBY, Smith2020BILBY} to perform PE for each event and recover the PE samples. The PE prior is assumed to be the aligned-spin prior, uniform in detector-frame masses and a power-law in luminosity distance.

To isolate the $35\,\Msun$ peak, we select events for which the PE posterior satisfies $m_1>20\,\Msun$ and $3\,\Msun<m_2<50\,\Msun$ with probability greater than $0.5$. The rest of the procedure is the same as described in Section \ref{sec:Method}. As shown in Fig.~\ref{fig: inj rec}, we successfully recover the injected values using Eq.~\eqref{eq:likelihood}.

\newcommand{\newblock}{}
\bibliographystyle{aasjournal}
\bibliography{main}
\end{document}